\documentclass[a4paper,11pt,fleqn]{article}

\usepackage{jheppub}
\usepackage{cancel}
\usepackage{jheppub}
\usepackage{subfig}
\usepackage{amsmath}

\usepackage{amssymb}
\usepackage{latexsym}
\usepackage{wasysym}
\usepackage{pstricks}

\newcommand{\bfl}{\begin{flalign}}

\newcommand{\defas}{\mathrel{\mathop:}=}
\newcommand{\mzv}[2][]{\zeta^{#1}_{#2 }}

\newcommand{\oz}{\overline{\zeta}}

\newlength{\myl}
\newcommand{\msbar}{\overline{\mbox{MS}}}

\newcommand{\bga}{\begin{gather}}

\newcommand{\ega}{\end{gather}}

\newcommand{\bal}{\begin{align}}

\newcommand{\eal}{\end{align}}

\newcommand{\z}{\zeta}

\newcommand{\ice}[1]{{}}

\newcommand{\EQN}[1]{\hspace{3mm}\fbox{\fbox{$#1$}} \label{#1}}

\newcommand{\ed}{\end{document}}

\newcommand{\prd}{\partial}
\newcommand{\ep}{\epsilon}
\newcommand{\beq}{\begin{equation}}
\newcommand{\eeq}{\end{equation}}
\newcommand{\bea}{\begin{eqnarray}}
\newcommand{\eea}{\end{eqnarray}}

\newcommand{\ba}{\begin{array}}
\newcommand{\ea}{\end{array}}

\newcommand{\g}{\gamma}

\newcommand{\be}{\beta}
\newcommand{\bc}{\begin{center}}
\newcommand{\ec}{\end{center}}

\newcommand{\re}[1]{(\ref{#1})}

\newcommand{\ovl}[1]{\overline{#1}}

\newcommand{\unb}[1]{\underbrace{#1}}

\catcode`\@=11
\def\slash{\mathpalette\make@slash}
\def\make@slash#1#2{\setbox\z@\hbox{$#1#2$}%
  \hbox to 0pt{\hss$#1/$\hss\kern-\wd0}\box0}
\catcode`\@=12 % @ signs are no longer letters

\def\bbuildrel#1_#2^#3%
{\mathrel{\mathop{\kern 0pt#1}\limits_{#2}^{#3}}}

\newcommand{\nnb}{\nonumber}

\newcommand{\MSbar}{\ensuremath{\overline{\text{MS}}}}
\def\beq{\begin{equation}}
\def\eeq{\end{equation}}
\def\bea{\begin{eqnarray}}
\def\eea{\end{eqnarray}}
\def\bq{\begin{quote}}
\def\eq{\end{quote}}

\def\nnb{\nonumber}

\def\nnb{\nonumber}

\def\ba{\begin{array}}
\def\ea{\end{array}}

\def\bbuildrel#1_#2^#3%
{\mathrel{\mathop{\kern 0pt#1}\limits_{#2}^{#3}}}

\newcommand{\lQ}{\ell_\mu}

\makeatletter
\newcommand\underparen[1]{%
  \mathop{%
    \vtop{
      \m@th
      \ialign{%
        ##\crcr
        $\hfil\displaystyle{#1}\hfil$\crcr
        \noalign{\kern3\p@\nointerlineskip}%
        \upparenfill\crcr
      }%
    }%
  }\limits
}
\newcommand\upparenfill{%
  $\m@th\setbox\z@\hbox{$\braceld$}%
  \bracelu\leaders\vrule \@height\ht\z@ \@depth\z@\hfill\braceru$%
}
\makeatother

\ice{ example: \underparen{\hat{\zeta}_3}_{L=6}}

\usepackage{mathtools} %
\ice{
\providecommand\given{}
 \DeclarePairedDelimiterXPP\EV[1]{E}(){}{
 \renewcommand\given{\nonscript\,\delimsize\vert\nonscript\,}
 #1}
 \DeclarePairedDelimiterXPP\Var[1]{V}(){}{
 \renewcommand\given{\nonscript\,\delimsize\vert\nonscript\,}
 #1}
}
\ice{ example: 

\begin{align}
  V(\varepsilon)&=\overbracket[0.4pt]{ E_{X}(\Var{\varepsilon \given X})}^\text{expected value}
+V_{X}(\smash[b]{\underbracket[0.6pt]{\EV{\varepsilon \given X}}_\text{$=0$}})\notag
\\ & =E_{X}(\Var{\varepsilon \given X})\notag \\
&=E(\sigma^{2}_{Y\vert X})
\end{align}

}

\title{
{
\boldmath
\!\!\!\!\!\!\!\!
Transcendental structure of  multiloop    massless correlators and 
anomalous dimensions
}
}
\author[a]{P.~A.~Baikov,}
\author[b,c]{K. G. Chetyrkin}

\affiliation[a]{
Skobeltsyn Institute of Nuclear Physics, Lomonosov Moscow State University, 
1(2), Leninskie gory, Moscow  119991, Russian Federation
}
\affiliation[b]{Institut f\"ur Theoretische Teilchenphysik, Karlsruhe
  Institute of Technology (KIT), Wolfgang-Gaede-Stra\ss{}e 1, 726128 Karlsruhe, Germany}

\affiliation[c]{
II Institut f\"ur Theoretische Physik,
Universit\"at  Hamburg, Luruper Chaussee 149, 22761 Hamburg, Germany
}

\emailAdd{baikov@theory.sinp.msu.ru}
\emailAdd{Konstantin.Chetyrkin@kit.edu}

\abstract{

 We give a short account of recent advances in our understanding of
  the $\pi$-dependent terms in massless (Euclidean) 2-point functions as well
  as in generic anomalous dimensions (ADs) and $\beta$-functions.  We extend the
  considerations of \cite{Baikov:2018wgs} by two more loops, that is for the
  case of 6- and 7-loop correlators and 7- and 8-loop renormalization group
  (RG) functions. Our predictions for the ($\pi$-dependent terms) of the
  7-loop RG functions   for the case
  of the $O(n)$ $\phi^4$ theory are in full agreement with the recent results
  from \cite{Schnetz:2016fhy}. All available 7- and 8-loop results  for QCD and the scalar
  $O(n)$
  $\varphi^4$ theory obtained within the large $N_f$ approach to the quantum
  field theory (see, e.g. \cite{Gracey:2018ame}) \ice{(for the loop numbers not exceeding 8)}
  are also in full agreement with our results.
}

\preprint{TTP19-026}


\renewcommand{\EQN}[1]{\label{#1}}

\begin{document}

%\today{}

\maketitle

\section{Introduction \label{sec:intro}}

Since the seminal calculation of the Adler function at order
$\alpha_s^3$ \cite{Gorishnii:1991vf} it has been known that
p-functions in QCD demonstrate striking regularities in \ice{as for}
terms proportional to $\pi^{2 n}$ (or, equivalently, even
zetas\footnote{As is well known every even power $2n$ of $\pi$ is
  uniquely related to the  corresponding Euler $\zeta$-function

$
  \zeta_{2n} \equiv  \sum_{i > 0} \frac{1}{i^{2n}}, \  \  \mbox{according  to a rule}
  \   \
  \zeta_{2n} = r(n)\, \pi^{2\,n}
{}, 
$
with $r(n)$ being  a (known) rational  number  \cite{MR360116}.},
with $n$ being positive integer.
Indeed, it was demonstrated in \cite{Gorishnii:1991vf} for the first time a mysterious complete
cancellation of all contributions proportional to
$\zeta_4 \equiv \frac{\pi^4}{90}$ (which generically appear in separate
diagrams) while odd zetas terms (that is those proportional to
$\zeta_3$ and $\zeta_5$ in the case under consideration) do survive  and show up in the final
result. 
Here by  p-functions we understand ($\msbar$-renormalized) Euclidean 
Green  functions\footnote{Like quark-quark-qluon vertex in QCD with
the external gluon line carrying no momentum.}
or 2-point correlators or even some combination thereof,
expressible in terms of massless propagator-like Feynman integrals
(to be named p-integrals below). 

\ice{Note that on general grounds the ${\cal O}(\alpha_s^3)$
Adler function
}
  
Since then it has been noted many times that {\em all} physical
(that is scale-invariant) 
p-functions are indeed free from even zetas at order $\alpha_s^4$
(like corrections to the Bjorken (polarized) DIS sum rule) and {\em
  some} of them---like the Adler function---even at the next, in fact,
5-loop, $\alpha_e \alpha_s^4$ order \cite{Baikov:2010je}. On the
other hand, the first appearance of $\zeta_4$ in a one-scale physical
quantity has been demonstrated in \cite{Baikov:2017ujl} for the case
of the 5-loop scalar correlator.

It should be stressed that the limitation to QCD p-functions in the
above discussion is {\em essential}. In general case  scale-invariant 
p-functions    do depend on even zetas already at 4 loops (see eq.~(11.8) in 
\cite{Baikov:2018wgs}). 
\ice{
For instance, for the case of a
much simpler the scalar $O(n)$ $\phi^4$ model the scale-invariant
p-functions do depend on even zetas already at 4 loops \cite{}.
}

To describe  these regularities   more precisely we need  to
introduce a few notations and conventions.
\ice{
(In what  follows we limit ourselves
by the case of QCD considered in the Landau gauge).
}
Let 
\beq
F_n(a,\lQ)  = 1  +\sum_{ 1\le i \le  n }^{0 \le j \le i} g_{i,j} \, (\ell_{\mu})^j \, a^i
\EQN{G:def}
{}
\eeq
be a (renormalized) p-function in a one-charge theory with the coupling constant  $a$\footnote{We implicitly assume that
the coupling constant  $a$ counts  loops. }.   
Here  $Q$ is  an (Euclidean) external momentum and    $\lQ= \ln \frac{\mu^2}{Q^2}$.
The integer $n$ stands for the (maximal)
power  of $a$ appearing in the p-integrals  contributing to
$F_n$. In  the  case of one-charge gauge  theory
and gauge non-invariant $F$ we will always assume the case of the Landau gauge. In particularly
all our  generic considerations in this paper are relevant for QCD p-functions with  
$a=\frac{\alpha_s(\mu)}{4\,\pi}$.

The $F$ without $n$ will stand as a shortcut for a formal series $F_\infty$.
In terms of bare quantities\footnote{
  We assume  the use of  the dimensional regularization  with the space-time dimension
$D=4-2\,\ep$.} 
\beq
F = Z\, F_B(a_B,\lQ), \hspace{2cm} Z  = 1 +\sum_{ i \ge 1}^{1 \le j \le i} 
Z_{i,j} \, \frac{a^i}{\ep^j}
%\ \ \mbox{with} \ \ \  
\EQN{G:GB}
{},
\eeq
with   the bare coupling  constant, the corresponding renormalization constant (RC) and AD being
\beq
a_B = \mu^{2\ep} Z_a \,a , \hspace{2cm}  Z_a   = 1 +\sum_{  i \ge 1 }^{1 \le j \le i} 
\Bigl(Z_a\Bigr)_{i,j} \, \frac{a^i}{\ep^j}
{},
\eeq
\ice{
Here $\ep$ is the standard parameter of dimensional regularization 
related with the running space-time dimension $D$ via $D=4-2\,\ep$.
The evolution equation for F reads:
}
\beq
\Bigl(\frac{\prd}{\prd \lQ}\, + \be\,a\, \frac{\prd}{\prd a}\Bigr) F = \g\, F
\EQN{G:RG:evol}
{},
\eeq
\beq
\g(a) =  \sum_{ i \ge 1 } \g_i \, a^i, \ \    \g_i = -i Z_{i,1}  
\EQN{gama:def}
{}.
\eeq
The coefficients of the $\beta$-function $\beta_i$ are related to  $Z_a$ in the 
standard way:
\beq
\beta_i = i \left(Z_a\right)_{i,1} 
\EQN{beta:coef:def}
{}.
\eeq
A p-function $F$ is called scale-independent if the corresponding AD $\gamma \equiv 0$.  
If $\gamma \not=0$ then one  can always construct a scale-invariant object from
 $F$ and $\g$, namely\ice{ 
The  trick was first applied for constructing a $\pi$-free version of a
 2-point  correlator in \cite{Vermaseren:1997fq}.
The same quantity as defined in \re{dG1} was discussed in  \cite{Kubo:1983ws} under the name of the
``Renormalization Scheme invariant anomalous dimension''. 
In DIS similar objects are called "physical anomalous
dimensions", see, e.g. \cite{Davies:2017hyl}. 
}:
\beq
F^{\mathrm si}_{n+1}(a,\lQ) = \frac{ \prd }{\prd \lQ} \left(\ln F\right)_{n+1}
\equiv \Biggl(\frac{\left( \g(a) - \beta(a) a \frac{\prd}{\prd a}\right)\,F_n}{F_n}\Biggr)_{n+1}
{}.
\EQN{dG1}
\eeq         
Note that $F^{\mathrm si}_{n+1}(a,\lQ)$ starts from the first power of the coupling
constant $a$ and is formally   composed from ${\cal O}(\alpha_s^{n+1})$ Feynman  diagrams.
In the same time  it can be completely restored from $F_n$ and the  $(n+1)$-loop AD $\g$.

If not otherwise stated we will assume  the so-called $G$-scheme for
renormalization \cite{Chetyrkin:1980pr}. The scheme is natural for massless propagators.  All
ADs, $\beta$-functions and $Z$-factors are identical in
\MSbar- and G-schemes.  For (finite) renormalized functions there exists a
simple conversion rule. Namely, in order to switch from an  $G$-renormalized
quantity to the one in the \MSbar-scheme one should make the following
replacement in the former: $\ln{\mu^2} \to \ln{\mu^2} +2$
($\mu$  is the renormalization scale, the limit of $\ep \to 0$ is understood).

An (incomplete) list of the  currently known  regularities\footnote{
For discussion of particular examples of  $\pi$-dependent contributions 
into various  QCD p-functions we refer  to works
 \cite{Jamin:2017mul,Davies:2017hyl,Chetyrkin:2017bjc,Ruijl:2018poj,Herzog:2018kwj}.}
 includes the  following cases.
\ice{
(for
gauge-dependent p-functions we assume  the use of the  Landau gauge). 
}
\begin{enumerate}
\item  Scale-independent QCD  p-functions $F_n$ and $F^{\mathrm si}_{n}$ with $n \le 4$  are  free 
from $\pi$-dependent terms.
\item  Scale-independent QCD p-functions $F^{\mathrm si}_5$ are free from $\pi^6$ and $\pi^2$ 
but do depend on $\pi^4$.
\item The QCD $\beta$-function starts  to depend on $\pi$ at  5 loops  only 
  \cite{Baikov:2016tgj,Herzog:2017ohr,Luthe:2017ttg}
  via $\z_4$.
In addition, there  exits a remarkable identity \cite{Baikov:2018wgs}
\[
\beta_5^{\z_4} = \frac{9}{8}  \beta_1^{(1)}\, \beta_4^{\z_3}, \    \  \ 
\mbox{with} \  \  \ F ^{\z_i} = \lim_{\z_i \to 0} \frac{\prd}{\prd \z_i} F 
%\frac{\prd}{\prd \zeta_3}
{}.
%\label{z3z4:beta}
\]
\ice{
where 
the upper-script $\z_i$  means
$
%\beq
 F ^{\z_i} = \lim_{\z_i \to 0} \frac{\prd}{\prd \z_i} F 
\ \ \mbox{and (for future reference)} \ \  
F ^{\z_i \z_j} = \lim_{\z_i \to 0} \frac{\prd}{\prd \z_i} F^{\z_j} 
{}.
$
%\eeq
}
\item  If we change   the $\msbar$-renormalization scheme as follows:
\beq
a = \bar{a} \, ( 1 + c_1\, \bar{a} +c_2\,  \bar{a}^2 + c_3 \, \bar{a}^3 
+ \frac{1}{3} \, \frac{\beta_5}{\beta_1^{(1)}}\, \bar{a}^4 ) 
\EQN{new:scheme:1}
{},
\eeq
with $c_1, c_2$ and  $c_3$ being any rational numbers, 
then all known  QCD functions  $\hat{F}^{\, \mathrm si}_5(\bar{a},\lQ)$ and  the (5-loop) QCD  $\beta$-function
$\bar{\beta}(\bar{a})$ both loose any dependence on  $\pi$. This remarkable fact
 was discovered 
in \cite{Jamin:2017mul} and led to the renewed  interest in the  issue of even zeta values
in two-point correlators and related objects.

\item
It should be also noted that no terms proportional to the first or second powers of
$\pi$  do ever appear in {\em all  known} (not necessarily QCD!)   p-functions and even in separate p-integrals
at least at loop number $L$  less or equal 5.
This comes straightforwardly from the fact that  the corresponding   master p-integrals 
are free from  such terms. The latter   has been established by explicit analytic calculations for
$L=2,3$ \cite{Chetyrkin:1980pr}, $L=4$ \cite{Baikov:2010hf,Lee:2011jt,Panzer:2013cha} and
finally\ice{
Note for this case only a part of  5-loop master integrals  was explicitly computed.
However, there are strong generic  mathematical arguments in favor of absence of contributions
with weight one and two, that is $\pi$ and $\pi^2$  
in   p-integrals {\em at least with the proper choice of the basis  set of transcendental generators}
\cite{}. } at $L=5$ \cite{Georgoudis:2018olj}.
Note for the last case only a part of  5-loop master integrals was explicitly computed.
However, there are  generic  mathematical arguments in favor of absence of contributions
with weight one and two, that is $\pi$ and $\pi^2$  
in   p-integrals {\em at least with the proper choice of the basis set of transcendental generators}
\cite{Brown:2015fyf,Panzer:2016snt}. 
\ice{
Pavel! here I have  2 poosible formulations 
By  {\em proper choice} here we mean, essentially,  a requirement that
{\em finite} p-integrals should be expressible in terms of  {\em rational} combinations of transcendental generators
(see eq. \re{hj:def}).
}
By  {\em proper choice} here we mean, essentially,  a requirement that
transcendental generators should be expressible in terms of  
{\em rational} combinations of {\em finite} p-integrals \cite{Broadhurst:1997kc,Broadhurst2013},
without use of $\pi$ as a generator.

Our results below are  in full agreement with these arguments.

\end{enumerate}

It should be stressed that  eventually  every separate diagram contributing to 
$F_n$ and $F^{\mathrm si}_{n+1}$ contains the following set of irrational numbers:
$\z_3, \z_4, \z_5, \z_6$ and $\z_7$ for $n =4$,
$\z_3$, $\z_4$ and $\z_5$ for $n=3$ as illustrated in  Table \ref{tab1}.
Thus,   the regularities listed above are quite nontrivial and  for sure  can not be explained by pure
coincidence.

\ice{
Seemingly unrelated interesting observations  can be made for
RG-functions (that is anomalous dimensions and $\beta$-functions).
Indeed, the QCD $\beta$-function  is free from any irrationals at $L=3$ \cite{}.
At next loop, $L=4$,   it depends on the only one, $\zeta_3$ \cite{} while
the  5-loop result includes, in addition, some  terms proportional to $\zeta_4$ and $\zeta_5$.
On the other  hand,  
on general
grounds based on the explicit  results for the corresponding  master integrals the possible
irrational structures which can   appear at multiloop p-integrals  display some 
}

In this paper\footnote{A preliminary version of the present work
  (not including   the 8-loop case) was reported on the International Seminar ``Loops and Legs in
  Quantum Field Theory'' (LL2018) in St. Goar, Germany and published in
  \cite{Baikov:2018gap}.}  we first present a short discussion of recent
advances in studying  the structure of the $\pi$-dependent terms in
massless (Euclidean) 2-point functions as well as in generic anomalous
dimensions and $\beta$-functions.  Then we extend the considerations of
\cite{Baikov:2018wgs} by two more loops. Finally, we discuss remarkable
connections between $\ep$-expansion of 4-loop p-integrals and   the $D=4$  values of finite
5-,  6-, and 7-loop p-integrals. 

\section{Hatted representation:  general formulation and   its implications}

%\section{Hatted representation   of p-integrals}
%\section{Main results of \cite{Baikov:2018wgs}}

The full understanding and a generic proof of points 1--5 above have
been recently achieved in our work \cite{Baikov:2018wgs}.  The main
tool of the work was the so-called ``hatted'' representation of
transcendental objects contributing to a given set of p-integrals.
\ice{
For
clarity we will start from the 4-loop case, then continue with more
loops and more abstract definition of the very hatted
representatation. The latter will be very helpful later in considering
the cases of 5- 6- and, finally, 7-loop p-integrals.
}

\renewcommand{\xcancel}[1]{}

\begin{center}
\begin{table}
\begin{center}
\begin{tabular}{ |c|c||c|c| } 
\hline
 L  & p-integrals  & L+1 &  Z 
\\
\hline
%\hline  
0  &  rational                          &  1  & $\mbox{rational}/\ep$ \\
1  &  rational/$\ep$                          &  2  & $\mbox{rational}/\ep^2$ \\
2  & $\z_3$                             &  3  &   $\z_3/\ep$ \\
3  & $\z_3/\ep,\, \z_4,\, \z_5$                 &  4  &  $\z_3/\ep^2,\, \{\z_4, \z_5\}/\ep$ \\
4  & $\z_3/\ep^2,\, \{\z_4, \z_5\}/\ep,\, \z_3^2,\, \z_6,\, \z_7 \xcancel{\z_3\z_4}$   &  5  %
&   $\z_3/\ep^3,\, \{\z_4, \z_5\}/\ep^2,\, \{\z_3^2, \z_6, \z_7 \xcancel{\z_3\z_4}\}/\ep$  \\
%\hline
%%%%%%%%%%%%%%%%%%%%%%%%%%%%% tab2
%\hline  
5  & 
$\z_3/\ep^3, \{\z_4, \z_5\}/\ep^2, \{\z_3^2, \z_6, \z_7\}/\ep$, 
&  6  %
&   $\z_3/\ep^4, \{\z_4, \z_5\}/\ep^3, \{\z_3^2, \z_6, \z_7  \xcancel{\z_3 \z_4}\}/\ep^2$, \\
\phantom{5}  & 
$\z_3\z_4$, $\z_8,\, \z_3\z_5,\, \z_{5,3},\, \z_3^3,\, \z_9\ice{,\, \z_4\z_5,\, \z_3\z_6}$  
&  \phantom{6}  %
&  $\{  \z_3\z_4, \z_8,\, \z_3\z_5,\,  \z_{5,3},\, \z_3^3,\, \z_9   \}/\ep$ \\
  \hline
%%%%%%%%%%%%%%%%%%%%%%%%%%%%% 
\end{tabular}
\caption{
\EQN{tab1}
\small The structure of p-integrals  (expanded in $\ep$ up to and including the  constant $\ep^0$ part)
% with $\ep$ set to zero  after renormalization is performed
and  RCs  in dependence on the loop number $L$. The 
inverse power of $\ep$ stands for the {\em  maximal} one  in generic case; in particular cases it
might be less. 
}
\end{center}
\end{table}
\end{center}

%\subsection{Hatted representation   of 4-loop p-integrals}
%\subsection{Hatted representation:  5- and   and 6-loop  cases}

Let us  reformulate   the main results of \cite{Baikov:2018wgs} in an abstract form. 
We will call  the set  of all  L-loop p-integrals ${\cal P}_L$  a $\pi$-safe one if   the following is  true.

(i)  All p-integrals from the set can  be expressed in terms of $(M + 1)$ 
mutually independent  (and $\ep$-independent) transcendental generators 
\beq
{\cal T}=\{t_1,t_2, \dots , t_{M+1}\}\ \ \mbox{with} \ t_{M+1} = \pi
{}.
\eeq
This means that any p-integral $F(\ep)$ from    ${\cal P}_L$ can be {uniquely}\footnote{
We assume that $F(\ep, {t}_1,{t}_2, \dots , \pi)$ does not  contain  terms  proportional to $\ep^n$
with $n \ge 1$.} presented  as follows 
\beq F(\ep) = F(\ep, {t}_1,{t}_2,\dots , t_{M}, \pi) + {\cal{ O}}(\ep)
\EQN{Fep}
{},
\eeq
\ice{
\beq
F(\ep) = F(\ep, \hat{t}_1,\hat{t}_2, \dots ,\hat{t}_M, \pi) + {\cal{ O}}(\ep)
{},
\eeq
}
\noindent
where by $F(\ep)$ we understand  the {\em exact}  value of  the p-integral $F$
while  the combination    $ \ep^L\,F(\ep, {t}_1,{t}_2, \dots ,{t}_M, \pi)$    should be  
a rational polynomial\footnote{
That is a polynomial having rational coefficients.}    in $\ep,t_1 \dots , t_{M},\pi$.
Every such polynomial is a sum of monomials $T_{\alpha}$  of the generic form
\beq
\sum_{\alpha} r_{\alpha}T_{\alpha}, \ \  T_{\alpha}=\,\ep^n\, \prod_{i=1,M+1}  t_i^{n_i} 
\EQN{T:def}
{},
\eeq
with $n \le L$, $n_i$ and $r_{\alpha}$   being some non-negative integers and  rational numbers respectively.   
A monomial
$T_\alpha$ will be called  {\em $\pi$-dependent} and denoted as $T_{\pi,\alpha}$ if $n_{M+1} > 0$. Note that a generator
$t_i$ with $i \le M$  may still   include explicitly the constant  $\pi$ in its definition, see below.

\ice{ 
while  $\ep^L F(\ep, \hat{t}_1,\hat{t}_2, \dots ,\hat{t}_M, 0,0,  \dots , 0)$ is a polynomial in $t_i$ with coefficients being finite polynomials  in $\ep$ over the field of the rational  numbers.  
}
%The difference between f and  F   
\
(ii)
For every $t_i$ with $i \le M$  let us define  its hatted counterpart as follows: 
\beq
\hat{t}_i =  t_i + \ep\!\!\! \  \sum_{\alpha}   h_{i\alpha}(\ep)\,  \,\,T_{\pi,\alpha}  
\EQN{hj:def}
{}, 
\eeq
with $\{h_{i\alpha}\}$ being rational polynomials in $\ep$
and $T_{\pi,\alpha}$  are {\em all}  $\pi$-dependent monomials as defined in \re{T:def}. 
Then there should exist  a choice of both  a basis ${\cal T}$ and  polynomials $\{h_{i\alpha}\}$ such that for every  
L-loop p-integral  $F(\ep, t_i)$ the following equality holds:
\beq
F(\ep, t_1,t_2,\dots ,{t}_M, \pi) = F(\ep, \hat{t}_1,\hat{t}_2, \dots , \hat{t}_M, 0) + {\cal{ O}}(\ep)
\EQN{hat:master}
{}.
\eeq
\ice{
We will call a (renormalized) function $F$ and/or an AD $\gamma$ {\em $\pi$-free} (both are assumed to be they
belong to a ring formed by generators $\{t_i,\ i= 1, \dots , M\}$ over the
field of rational numbers.
}
We will  call  {\em $\pi$-free } 
any rational polynomial (with possibly $\ep$-dependent coefficients) in  $\{t_i,\ i= 1, \dots , M\}$.

As  we  will discuss below the sets ${\cal P}_i$ with $i=3,4,5$ are for 
sure  $\pi$-safe (well, for $L=5$ almost)  while 
${\cal P}_6$  highly likely shares the property. For the case of  ${\cal P}_7$  the situation is
more complicated (but still not hopeless!) as discussed in Section \ref{pi12}. 
In what follows we will always assume that
every (renormalized)  L-loop p-function as well as (L+1)-loop  $\beta$-functions and 
anomalous dimensions  are all  expressed in terms of the  generators $t_1,t_2,\dots , t_{M+1}$.

Moreover,  for any   polynomial $P(t_1,t_2,\dots , \pi)$  we define its {\em hatted}
version as
\[
\hat{P}(\hat{t}_1,\hat{t}_2,\dots , \hat{t}_{M}) \defas P(\hat{t}_1,\hat{t}_2,\dots , \hat{t}_{M}, 0) 
{}.
\]
Let 
$F_L$
is a (renormalized, with $\ep$ set to zero) p-function, $\gamma_L$ and $\beta_L$ 
are  the  corresponding anomalous dimension and  the $\beta$-function (all taken in the $L$-loop
approximation).
The following statements have been proved in \cite{Baikov:2018wgs} {\em under the 
condition that the  set ${\cal P}_L$ is $\pi$-safe and that both the  set ${\cal T}$ and 
the polynomials $\{h_{i\alpha}(\ep)\}$ are fixed}.

\vglue 0.5cm
\noindent
{\bf 1.  No-$\pi$ Theorem } 
\\
(a)
$F_L$
is $\pi$-free in any (massless) renormalization scheme for which  corresponding   $\beta$-function 
and AD $\g$ are both $\pi$-free at least at the level of $L+1$  loops.  
\\
(b)
The scale-invariant 
combination  $F_{L+1}^{\mathrm si}$ is $\pi$-free in  
any (massless) renormalization scheme provided  the   $\beta$-function
is   $\pi$-independent  at least at the level of $L+1$ loops. 
 
\vglue 0.2cm
%\vspace{6mm}

\vglue 0.5cm
\noindent
{\bf 2. $\pi$-dependence of L-loop p-functions  }
\\
If $F_L$ is renormalized in $\msbar$-scheme, then all its $\pi$-dependent
contributions can be expressed
in terms of $\hat{F}_{L}|_{\ep=0}$,
$\hat{\beta}_{L-1}|_{\ep=0}$ and $\hat{\gamma}_{L}|_{\ep=0}$.

\vglue 0.5cm
\noindent
{\bf 3. $\pi$-dependence of L-loop $\beta$-functions and AD  }  
\\
If $\beta_L$ and $\gamma_L$ are given in the $\msbar$-scheme, then all their
$\pi$-dependent contributions can be expressed in terms of
$\hat{\beta}_{L-1}|_{\ep=0}$ and $\hat{\beta}_{L-1}|_{\ep=0}$,
$\hat{\gamma}_{L-1}|_{\ep=0}$ correspondingly.

\section{$\pi$-structure of 3,4,5 and 6-loop p-integrals}

A hatted representation of  p-integrals is  known for  loop numbers $L=3$ \cite{Broadhurst:1999xk},
$L=$4 \cite{Baikov:2010hf}
and 
$L=5$ \cite{Georgoudis:2018olj}. 
In all three cases it was constructed by looking for such a basis ${\cal T}$ as
well as  polynomials $h_{i\alpha}(\ep)$ (see eq. \re{hj:def})  that
eq. \re{hat:master} would be valid for sufficiently  large subset of $ {\cal P}_L$.
\ice{
In all three cases it was constructed by  tuning  polynomials $h_\alpha(\ep)$
(see eq. \re{hj:def}) so that  eq. \re{hat:master} would be valid for all corresponding  master integrals.
}

Let us consider the next-loop level, that is ${\cal P}_6$.
In principle, \ice{for a loop number $L$}  the strategy requires the
knowledge of all (or almost all) L-loop master integrals.  On the other hand,
if we {\em assume} the $\pi$-safeness of the set ${\cal P}_6$ we could try to
fix polynomials $h_{i\alpha}(\ep)$ by considering some limited subset of ${\cal P}_6$.

Actually, we  do have at  our disposal a subset of ${\cal P}_6$ due to work \cite{Lee:2011jt}
where all 4-loop master integrals  have been computed up to the transcendental  weight  12 in their 
$\ep$-expansion.
Every particular 4-loop p-integral  divided by $\ep^n$ can be considered as a $(4+n)$-loop p-integral.
The collection of such (4+n)-loop p-integrals form a subset of ${\cal P}_{4+n}$ which we will
refer to as ${\cal P}_{4}/\ep^n$.
We have  tried  this subset for  $n=1$ and $2$.

Our results are given  below.
(To make resulting formulas shorter we  use even    zetas
\ice{
 Zeta[4]                                                                                                                          
          4
        Pi
Out[1]= ---
        90

In[2]:= Zeta[6]//InputForm    
                              
Out[2]//InputForm= Pi^6/945   

In[3]:= Zeta[8]//InputForm    
                              
Out[3]//InputForm= Pi^8/9450  

In[4]:= Zeta[10]//InputForm   
                              
Out[4]//InputForm= Pi^10/93555
}
$\z_2=\pi^2/6$, $\z_4 =\pi^4/90,$ $ \z_6 =  \pi^6/945, \z_8 = \pi^8/9450$ and $\z_{10}=  \pi^{10}/93555$
instead of  the corresponding  even powers of $\pi$).

\newcommand{\Unb}[2]{\underbracket[0.65pt][0.9mm]{\hspace{2.mm} #1 \hspace{1.mm}}_{L=#2}}

\begin{flalign}
  \Unb{\hat{\zeta}_3}{6} \defas \unb{\fbox{$\zeta_3$} + \frac{3 \epsilon}{2} \zeta_4}_{L=3} \qquad
  \unb{- \frac{5 \epsilon^3}{2} \zeta_6}_{\delta (L=4)}
\qquad  \unb{+\frac{21 \epsilon^5}{2} \zeta_8}_{\delta (L=5)} 
\qquad
\unb{-\frac{153\ep^7}{2} \z_{10}}_{\delta(L=6)}
{},
&&
\EQN{hz3}
\end{flalign}
%%%%%%%%%%%%%%%%%%%%%%%%%%%%%%%%%%%%%%%%%%%%%%%%%%%%%%%%%%%%%%%%%%%%%%% 
%
\vspace{-3mm}

\begin{flalign}
\Unb{\hat{\zeta}_5}{6}  \defas \unb{\fbox{$\zeta_5$}}_{L=3}   \qquad \unb{\ + \frac{5 \epsilon}{2} \zeta_6}_{\delta L=4} \qquad 
\unb{- \frac{35 \epsilon^3}{4} \zeta_8}_{\delta (L=5)} 
\qquad \unb{ +63 \ep^5 \z_{10} }_{\delta(L=6)}
{}, &&
\EQN{hz5}
\end{flalign}
\vspace{-3mm}
\bfl
\Unb{\hat{\zeta}_7}{6}  \defas \unb{\fbox{$\zeta_7$}}_{L=4} \qquad 
\unb{+ \frac{7 \epsilon}{2} \zeta_8}_{\delta(L=5)}
\qquad \unb{- 21 \ep^3 \z_{10}}_{\delta(L=6)}  
{},&&
\EQN{hz7}
\end{flalign}
\vspace{-6mm}
\ice{
\bfl 
\Unb{\hat{\varphi}}{6} 
\defas \unb{\fbox{$\varphi$}
 -3\epsilon\, \zeta_4 \,\zeta_5 + \frac{5 \epsilon}{2} \zeta_3\, \zeta_6 }_{L=5}
\qquad
\unb{
- \frac{24\,\ep^2}{47}\z_{10} + \ep^3\,( - \frac{35}{4}\z_{3}\z_{8} + 5\z_{5}\z_{6})
}_{\delta(L=6)}
{},
 &&
\EQN{hphi}
\end{flalign}
}

\begin{flalign} \Unb{\hat{ \z}_{5,3} }{6}    = 
\unb{
  \fbox{ $\z_{5,3}  
    - \frac{29}{12}  \zeta_{8} 
       $} - \frac{15\ep}{2}\zeta_{4}   \zeta_{5}
}_{L=5} \,  
\unb{
       - \frac{2905\ep^2}{376}  \zeta_{10} 
       + \frac{25\ep^3}{2}  \zeta_{5}   \zeta_{6}
     }_{\delta(L=6)}
\ice{       + ?  \zeta_{12} 
       - \frac{105\ep^5}{2}  \zeta_{5}   \zeta_{8} 
}
{},
&&\EQN{hz53}
 \end{flalign} \vspace{\myl}
 \vspace{1mm}

\bfl            
\Unb{\hat{\z_9}}{6} \defas \unb{\fbox{$\zeta_9$}}_{L=5} \qquad \unb{+ \frac{9}{2} \ep\,  \z_{10}}_{\delta(L=6)}
{},
&&
\EQN{hz9}
\end{flalign}
\vspace{-3mm}
\bfl
 \Unb{\hat{\z}_{7,3}}{6}\defas \unb{\fbox{$\z_{7,3}  - \frac{793}{94}\z_{10}$}   + 3\ep ( - 7\z_{4}\z_{7} - 5\z_{5}\z_{6})}_{L=6}
{},
 &&
\EQN{hz73}
\end{flalign}
\vspace{-3mm}
\bfl
\Unb{\hat{\z}_{11}}{6}\defas \unb{ \fbox{$\z_{11}$}
}_{L=6}
{},
 &&
\EQN{hz11}
\end{flalign}
\vspace{-3mm}
\bfl
\Unb{\hat{\z}_{5,3,3}}{6} \defas \unb{\fbox{$\z_{5,3,3}  + 45\z_{2}\z_{9} + 3\z_{4}\z_{7} -\frac{5}{2}\z_{5}\z_{6}$ }}_{L=6}
{}.
 &&
\EQN{hz5z3z3}
\end{flalign}
Here 
multiple zeta values are defined  as \cite{Blumlein:2009cf}
\beq
\mzv{n_1,n_2} \defas \sum_{i > j > 0 } \frac{1}{i^{n_1} j^{n_2}}, \ \
\mzv{n_1,n_2,n_3} \defas \sum_{i > j >  k > 0 } \frac{1}{i^{n_1} j^{n_2} k^{n_3}  }
\EQN{mzv}
{}.
\eeq
Some comments on these eqs. are  in order.

\begin{itemize}

\item The boxed entries form a set of $\pi$-independent (by definition!) generators  for
the cases of $L=3$ (eqs. (\ref{hz3}, \ref{hz5}), $L=4$ (eqs. (\ref{hz3}---\ref{hz7})),  
$L=5$ (eqs. (\ref{hz3}--\ref{hz9})) and  $L=6$
(eqs. (\ref{hz3}---\ref{hz5z3z3})).
In what follows  we will use for the boxed combinations in eqs. (\ref{hz3} -\ref{hz5z3z3})
the notation
\beq
\ovl{\zeta}_{n_1,n_2, \dots}  \defas  \hat{\zeta}_{n_1,n_2,\dots}|_{\ep=0}.
\EQN{barmvz}
%\ovl{\zeta}_{n_1,n_2,n_3} & \defas  \hat{\zeta}_{n_1,n_2}|{\ep=0}.
\eeq

\item

  There is no terms proportional to single or second  powers of $\pi$  {\em outside}
  boxed combinations in relations  (\ref{hz3}--\ref{hz5z3z3}). This  fact directly
  leads to the absence of such terms in  the (renormalized) 6-loop p-integrals
  and generic ADs  and $\beta$-functions with  the loop number not
  exceeding 7. Later we  will see that the same is true for 7-loop p-integrals and
  8-loop RG functions (assuming the conservative scenario as described in Section \ref{pi12}).

\ice{   
\item For $L=5$ we recover  the
hatted representation  
for the set ${\cal P}_5$ first found  in \cite{Georgoudis:2018olj}. The latter
coincides with eqs. (\ref{hz3}-\ref{hz7}) and (\ref{hz9}) while instead of
(\ref{hz53}) the authors of \cite{Georgoudis:2018olj} suggest
\beq
\hat{\varphi} 
                \defas \varphi -3\epsilon\, \zeta_4 \,\zeta_5 + \frac{5 \epsilon}{2} \zeta_3\, \zeta_6
\EQN{phi}
{},
\eeq
with
\ice{$\varphi=\mzv{6,2}- \mzv{3,5} \approx -0.1868414$ }
\beq
\varphi \defas \frac{3}{5}\, \z_{5,3} + \z_3\, \z_5 -\frac{29}{20} \,\z_8 
= \mzv{6,2}- \mzv{3,5}
\EQN{varphi}
{}.
\eeq
%\ed
Note that in order to fit eq.~\re{hj:def} one should redefine the  hatted object which corresponds to $\zeta_{5,3}$.
Our $\hat{\zeta}_{5,3}$ is a linear combination of $\hat{\zeta}_{3,5}$ and
$\hat{\zeta}_3 \,\hat{\zeta}_5$ from \cite{Georgoudis:2018olj}. 
}
\item
For $L=5$ we recover  the
hatted representation  
for the set ${\cal P}_5$ first found  in \cite{Georgoudis:2018olj}. The latter
coincides with eqs. (\ref{hz3}-\ref{hz7}) and (\ref{hz9}) while instead of
(\ref{hz53}) the authors of \cite{Georgoudis:2018olj} suggest
\beq
\hat{\z}_{3,5} 
                \defas \varphi -3\epsilon\, \zeta_4 \,\zeta_5 + \frac{5 \epsilon}{2} \zeta_3\, \zeta_6
\EQN{phi}
{},
\eeq
with
\ice{$\varphi=\mzv{6,2}- \mzv{3,5} \approx -0.1868414$ }
\beq
\varphi \defas \frac{3}{5}\, \z_{5,3} + \z_3\, \z_5 -\frac{29}{20} \,\z_8 
= \mzv{6,2}- \mzv{3,5}
\EQN{varphi}
{}.
\eeq
%\ed
\ice{ Note that in order to fit eq.~\re{hj:def} one should redefine
  the hatted object which corresponds to $\zeta_{5,3}$.  Our
  $\hat{\zeta}_{5,3}$ is a linear combination of $\hat{\zeta}_{3,5}$
  and $\hat{\zeta}_3 \,\hat{\zeta}_5$ from \cite{Georgoudis:2018olj}.
}
Our $\hat{\zeta}_{5,3}$ (eq. \re{hz53}) is related (up to the
corresponding order of $\epsilon$) to $\hat{\zeta}_{3,5}$ from \cite{Georgoudis:2018olj}
 as
$\hat{\zeta}_{5,3}=\frac{5}{3}(\hat{\zeta}_{3,5}-\hat{\zeta}_3\,\hat{\zeta}_5)$.
The reason for this redefinition is that we want
every hatted zeta to be  equal to the corresponding unhatted zeta plus
terms  proportional
to explicit powers of $\pi^2$ at $\epsilon^0$ order as well.

\item We do not claim that the generators
  \beq \z_3,\z_5,\z_7, \oz_{5,3},
  \z_9,  \oz_{7,3}, {\z}_{11}, \oz_{5,3,3} \ \mbox{and} \ \ \pi
  \EQN{gens}
  \eeq
  are sufficient to present the pole and finite parts
  of every 6-loop p-integral. In fact, it is not true
  \cite{Broadhurst:1995km:improved,Panzer:2015ida,Schnetz:2016fhy}.  However we
  believe \ice{think} that it is safe to assume that all missing
  irrational constants can be associated with the values of some
  convergent 6-loop p-integrals at $\ep=0$.

\end{itemize}

%The last comment deserves some elaboration.

\section{$\pi$-dependence of 7-loop $\beta$-functions and AD}

Using the approach of \cite{Baikov:2018wgs} and the hatted representation of
the irrational generators \re{gens} as described by eqs. \re{hz3}-\re{hz5z3z3}
\ice{for the 6-loop case} we can straightforwardly predict the $\pi$-dependent
terms in the $\beta$-function and the anomalous dimensions in the case of {\em
  any} 1-charge minimally renormalized field model at the level of 7 loops.

Our results read (the
combination $F^{ t_{\alpha_1} t_{\alpha_2}\dots  t_{\alpha_n}}$ stands for
the coefficient of the monomial  $(t_{\alpha_1} t_{\alpha_2}\dots  t_{\alpha_n})$ in $F$; in addition,  by
$F^{(1)}$ we understand $F$ with every  generator $t_i$  from $\{t_1, t_2,  \dots , t_{M+1}\}$ set to zero).

\newcommand{\brk}{\nnb & \\ & \hspace{9mm}  }

      \begin{flalign}
 & \gamma_{4}^{ \z_{4}  } =  
           -
\frac{1}{2}\,    \gamma_1^{(1)}  \beta_{3}^{ \z_{3}  } +
\frac{3}{2}\,  \gamma_{3}^{ \z_{3}  }    \beta_1^{(1)} {},&\EQN{a4z4}
  \end{flalign} \vspace{\myl}

      \begin{flalign}
 & \gamma_{5}^{ \z_{4}  } =  
           -
\frac{3}{8}\,    \gamma_1^{(1)}  \beta_{4}^{ \z_{3}  }   -
\gamma_2^{(1)}  \beta_{3}^{ \z_{3}  } +
\frac{3}{2}\,  \gamma_{3}^{ \z_{3}  }    \beta_2^{(1)} +
\frac{3}{2}\,  \gamma_{4}^{ \z_{3}  }    \beta_1^{(1)} {},&\EQN{a5z4}
  \end{flalign} \vspace{\myl}

      \begin{flalign}
 & \gamma_{5}^{ \z_{6}  } =  
           -
\frac{5}{8}\,    \gamma_1^{(1)}  \beta_{4}^{ \z_{5}  } +
\frac{5}{2}\,    \beta_1^{(1)}  \gamma_{4}^{ \z_{5}  } {},&\EQN{a5z6}
  \end{flalign} \vspace{\myl}

      \begin{flalign}
 & \gamma_{5}^{ \z_{4} \z_{3}  } =  0 {},&\EQN{a5z4z3}
  \end{flalign} \vspace{\myl}

      \begin{flalign}
 & \gamma_{6}^{ \z_{4}  } =  
           -
\frac{3}{10}\,    \gamma_1^{(1)}  \beta_{5}^{ \z_{3}  }   -
\frac{3}{4}\,    \gamma_2^{(1)}  \beta_{4}^{ \z_{3}  }   -
\frac{3}{2}\,    \gamma_3^{(1)}  \beta_{3}^{ \z_{3}  } +
\frac{3}{2}\,  \gamma_{3}^{ \z_{3}  }    \beta_3^{(1)} +
\frac{3}{2}\,  \gamma_{4}^{ \z_{3}  }    \beta_2^{(1)} 
         +
\frac{3}{2}\,  \gamma_{5}^{ \z_{3}  }    \beta_1^{(1)} {},&\EQN{a6z4}
  \end{flalign} \vspace{\myl}

      \begin{flalign}
 & \gamma_{6}^{ \z_{6}  } =  
           -
\frac{1}{2}\,    \gamma_1^{(1)}  \beta_{5}^{ \z_{5}  }   -
\frac{5}{4}\,    \gamma_2^{(1)}  \beta_{4}^{ \z_{5}  } +
\frac{5}{2}\,    \beta_2^{(1)}  \gamma_{4}^{ \z_{5}  } 
\brk
+
\frac{5}{2}\,    \beta_1^{(1)}  \gamma_{5}^{ \z_{5}  } +
\frac{3}{2}\,  (     \beta_1^{(1)}  )^2   \beta_{3}^{ \z_{3}  } 
            \gamma_1^{(1)}   -
\frac{5}{2}\,  (     \beta_1^{(1)}  )^3   \gamma_{3}^{ \z_{3}  } {},&\EQN{a6z6}
  \end{flalign} \vspace{\myl}

      \begin{flalign}
 & \gamma_{6}^{ \z_{8}  } =  
           -
\frac{7}{10}\,    \gamma_1^{(1)}  \beta_{5}^{ \z_{7}  } +
\frac{7}{2}\,    \beta_1^{(1)}  \gamma_{5}^{ \z_{7}  } {},&\EQN{a6z8}
  \end{flalign} \vspace{\myl}

      \begin{flalign}
 & \gamma_{6}^{ \z_{4} \z_{3}  } =  
           -
\frac{3}{5}\,    \gamma_1^{(1)}  \beta_{5}^{ \z_{3}^2  } +
3    \beta_1^{(1)}  \gamma_{5}^{ \z_{3}^2  } {},&\EQN{a6z4z3}
  \end{flalign} \vspace{\myl}

      \begin{flalign}
 & \gamma_{6}^{ \z_{4} \z_{5}  } =  0 {},&\EQN{a6z4z5}
  \end{flalign} \vspace{\myl}

      \begin{flalign}
 & \gamma_{6}^{ \z_{6} \z_{3}  } =  0 {},&\EQN{a6z6z3}
  \end{flalign} \vspace{\myl}

      \begin{flalign}
 & \gamma_{7}^{ \z_{4}  } =  
           -
\frac{1}{4}\,    \gamma_1^{(1)}  \beta_{6}^{ \z_{3}  }   -
\frac{3}{5}\,    \gamma_2^{(1)}  \beta_{5}^{ \z_{3}  }   -
\frac{9}{8}\,    \gamma_3^{(1)}  \beta_{4}^{ \z_{3}  } +
\frac{3}{2}\,  \gamma_{3}^{ \z_{3}  }    \beta_4^{(1)}   -
2    \gamma_4^{(1)}  \beta_{3}^{ \z_{3}  } 
\brk
+
\frac{3}{2}\,  \gamma_{4}^{ \z_{3}  }    \beta_3^{(1)} +
\frac{3}{2}\,  \gamma_{5}^{ \z_{3}  }    \beta_2^{(1)} +
\frac{3}{2}\,  \gamma_{6}^{ \z_{3}  }    \beta_1^{(1)} {},&\EQN{a7z4}
  \end{flalign} \vspace{\myl}

      \begin{flalign}
 & \gamma_{7}^{ \z_{6}  } =  
           -
\frac{5}{12}\,    \gamma_1^{(1)}  \beta_{6}^{ \z_{5}  }   -
\gamma_2^{(1)}  \beta_{5}^{ \z_{5}  }   -
\frac{15}{8}\,    \gamma_3^{(1)}  \beta_{4}^{ \z_{5}  } +
\frac{5}{2}\,    \beta_3^{(1)}  \gamma_{4}^{ \z_{5}  } +
\frac{5}{2}\,    \beta_2^{(1)}  \gamma_{5}^{ \z_{5}  } 
\brk
+
\frac{5}{2}\,    \beta_1^{(1)}  \gamma_{6}^{ \z_{5}  } +
\frac{5}{2}\,    \beta_1^{(1)}    \beta_2^{(1)}  \beta_{3}^{ \z_{3}  }    \gamma_1^{(1)} +
\frac{5}{4}\,  (     \beta_1^{(1)}  )^2   \beta_{4}^{ \z_{3}  }    \gamma_1^{(1)}
\brk
+
3  (     \beta_1^{(1)}  )^2   \beta_{3}^{ \z_{3}  }    \gamma_2^{(1)}   -
\frac{15}{2}\, 
          (     \beta_1^{(1)}  )^2     \beta_2^{(1)}  \gamma_{3}^{ \z_{3}  }   -
\frac{5}{2}\,  (     \beta_1^{(1)}  )^3   \gamma_{4}^{ \z_{3}  } {},&\EQN{a7z6}
  \end{flalign} \vspace{\myl}

      \begin{flalign}
 & \gamma_{7}^{ \z_{8}  } =  
           -
\frac{7}{12}\,    \gamma_1^{(1)}  \beta_{6}^{ \z_{7}  }   -
\frac{7}{5}\,    \gamma_2^{(1)}  \beta_{5}^{ \z_{7}  } +
\frac{7}{12}\,  (   \beta_{3}^{ \z_{3}  }  )^2     \gamma_1^{(1)} +
\frac{7}{2}\,    \beta_2^{(1)}  \gamma_{5}^{ \z_{7}  } 
\brk
+
\frac{7}{2}\,    \beta_1^{(1)}  
         \gamma_{6}^{ \z_{7}  }   -
\frac{7}{8}\,    \beta_1^{(1)}    \gamma_1^{(1)}  \beta_{5}^{ \z_{3}^2  }   -
\frac{7}{8}\,    \beta_1^{(1)}  \beta_{3}^{ \z_{3}  }  \gamma_{3}^{ \z_{3}  } 
\brk
+
\frac{21}{8}\,  (     \beta_1^{(1)}  )^2   \gamma_{5}^{ \z_{3}^2  } +
\frac{35}{8}\,  
         (     \beta_1^{(1)}  )^2     \gamma_1^{(1)}  \beta_{4}^{ \z_{5}  }   -
\frac{35}{4}\,  (     \beta_1^{(1)}  )^3   \gamma_{4}^{ \z_{5}  } {},&\EQN{a7z8}
  \end{flalign} \vspace{\myl}

      \begin{flalign}
 & \gamma_{7}^{ \z_{10}  } =  
           -
\frac{3}{4}\,    \gamma_1^{(1)}  \beta_{6}^{ \z_{9}  } +
\frac{9}{2}\,    \beta_1^{(1)}  \gamma_{6}^{ \z_{9}  } {},&\EQN{a7z10}
  \end{flalign} \vspace{\myl}

      \begin{flalign}
 & \gamma_{7}^{ \z_{4} \z_{3}  } =  
           -
\frac{1}{2}\,    \gamma_1^{(1)}  \beta_{6}^{ \z_{3}^2  }   -
\frac{6}{5}\,    \gamma_2^{(1)}  \beta_{5}^{ \z_{3}^2  } +
\frac{3}{8}\,  \gamma_{3}^{ \z_{3}  }  \beta_{4}^{ \z_{3}  }   -
\frac{1}{2}\,  \gamma_{4}^{ \z_{3}  }  \beta_{3}^{ \z_{3}  } +
3    \beta_2^{(1)}  
         \gamma_{5}^{ \z_{3}^2  } +
3    \beta_1^{(1)}  \gamma_{6}^{ \z_{3}^2  } {},&\EQN{a7z4z3}
  \end{flalign} \vspace{\myl}

      \begin{flalign}
 & \gamma_{7}^{ \z_{4} \z_{5}  } =  
        \frac{5}{4}\,    \gamma_1^{(1)}  \beta_{6}^{ \ovl{\z}_{5,3}  }   -
\frac{1}{4}\,    \gamma_1^{(1)}  \beta_{6}^{ \z_{3} \z_{5}  } +
\frac{3}{2}\,  \gamma_{3}^{ \z_{3}  }  \beta_{4}^{ \z_{5}  }   -
2  \beta_{3}^{ \z_{3}  }  \gamma_{4}^{ \z_{5}  }   -
\frac{15}{2}\,    \beta_1^{(1)}  
         \gamma_{6}^{ \ovl{\z}_{5,3}  } +
\frac{3}{2}\,    \beta_1^{(1)}  \gamma_{6}^{ \z_{3} \z_{5}  } {},&\EQN{a7z4z5}
  \end{flalign} \vspace{\myl}

      \begin{flalign}
 & \gamma_{7}^{ \z_{4} \z_{7}  } =  0 {},&\EQN{a7z4z7}
  \end{flalign} \vspace{\myl}

      \begin{flalign}
 & \gamma_{7}^{ \z_{6} \z_{3}  } =  
           -
\frac{5}{12}\,    \gamma_1^{(1)}  \beta_{6}^{ \z_{3} \z_{5}  }   -
\frac{15}{8}\,  \gamma_{3}^{ \z_{3}  }  \beta_{4}^{ \z_{5}  } +
\frac{5}{2}\,  \beta_{3}^{ \z_{3}  }  \gamma_{4}^{ \z_{5}  } +
\frac{5}{2}\,    \beta_1^{(1)}  \gamma_{6}^{ \z_{3} \z_{5}  } {},&\EQN{a7z6z3}
  \end{flalign} \vspace{\myl}

      \begin{flalign}
 & \gamma_{7}^{ \z_{6} \z_{5}  } =  0 {},&\EQN{a7z6z5}
  \end{flalign} \vspace{\myl}

      \begin{flalign}
 & \gamma_{7}^{ \z_{8} \z_{3}  } =  0 {},&\EQN{a7z8z3}
  \end{flalign} \vspace{\myl}

      \begin{flalign}
 & \gamma_{7}^{ \z_{4} \z_{3}^2  } =  
           -
\frac{3}{4}\,    \gamma_1^{(1)}  \beta_{6}^{ \z_{3}^3  } +
\frac{9}{2}\,    \beta_1^{(1)}  \gamma_{6}^{ \z_{3}^3  } {}.&\EQN{a7z4z3z3}
  \end{flalign} 
%\vspace{\myl}

The results for $\pi$-dependent contributions to a $\beta$-function are
obtained from the above relations by a formal replacement of $\gamma$ by
$\beta$ in every term. For instance, the 6 and 7-loop $\pi$-dependent
contributions read:
   
      \begin{flalign}
 & \beta_{6}^{ \z_{4}  } =  
        \frac{6}{5}\,    \beta_1^{(1)}  \beta_{5}^{ \z_{3}  } +
\frac{3}{4}\,    \beta_2^{(1)}  \beta_{4}^{ \z_{3}  } {},&\EQN{b6z4}
  \end{flalign} \vspace{\myl}

      \begin{flalign}
 & \beta_{6}^{ \z_{6}  } =  
        2    \beta_1^{(1)}  \beta_{5}^{ \z_{5}  }   -
(     \beta_1^{(1)}  )^3   \beta_{3}^{ \z_{3}  } +
\frac{5}{4}\,    \beta_2^{(1)}  \beta_{4}^{ \z_{5}  } {},&\EQN{b6z6}
  \end{flalign} \vspace{\myl}

      \begin{flalign}
 & \beta_{6}^{ \z_{8}  } =  
        \frac{14}{5}\,    \beta_1^{(1)}  \beta_{5}^{ \z_{7}  } {},&\EQN{b6z8}
  \end{flalign} \vspace{\myl}

      \begin{flalign}
 & \beta_{6}^{ \z_{4} \z_{3}  } =  
        \frac{12}{5}\,    \beta_1^{(1)}  \beta_{5}^{ \z_{3}^2  } {},&\EQN{b6z4z3}
  \end{flalign} \vspace{\myl}

      \begin{flalign}
 & \beta_{6}^{ \z_{4} \z_{5}  } =  0 {},&\EQN{b6z4z5}
  \end{flalign} \vspace{\myl}

      \begin{flalign}
 & \beta_{6}^{ \z_{6} \z_{3}  } =  0 {},&\EQN{b6z6z3}
  \end{flalign} \vspace{\myl}

      \begin{flalign}
 & \beta_{7}^{ \z_{4}  } =  
        \frac{5}{4}\,    \beta_1^{(1)}  \beta_{6}^{ \z_{3}  } +
\frac{9}{10}\,    \beta_2^{(1)}  \beta_{5}^{ \z_{3}  } +
\frac{3}{8}\,    \beta_3^{(1)}  \beta_{4}^{ \z_{3}  }   -
\frac{1}{2}\,  \beta_{3}^{ \z_{3}  }    \beta_4^{(1)} {},&\EQN{b7z4}
  \end{flalign} \vspace{\myl}

      \begin{flalign}
 & \beta_{7}^{ \z_{6}  } =  
        \frac{25}{12}\,    \beta_1^{(1)}  \beta_{6}^{ \z_{5}  } +
\frac{3}{2}\,    \beta_2^{(1)}  \beta_{5}^{ \z_{5}  }   -
\frac{5}{4}\,  \beta_{4}^{ \z_{3}  }  (     \beta_1^{(1)}  )^3  +
\frac{5}{8}\,    \beta_3^{(1)}  \beta_{4}^{ \z_{5}  }   -
2  \beta_{3}^{ \z_{3}  }    \beta_2^{(1)}  
         (     \beta_1^{(1)}  )^2  {},&\EQN{b7z6}
  \end{flalign} \vspace{\myl}

      \begin{flalign}
 & \beta_{7}^{ \z_{8}  } =  
        \frac{35}{12}\,    \beta_1^{(1)}  \beta_{6}^{ \z_{7}  } +
\frac{7}{4}\,  (     \beta_1^{(1)}  )^2   \beta_{5}^{ \z_{3}^2  }   -
\frac{35}{8}\,  (     \beta_1^{(1)}  )^3   \beta_{4}^{ \z_{5}  } +
\frac{21}{10}\,    \beta_2^{(1)}  \beta_{5}^{ \z_{7}  }   -
\frac{7}{24}\,  
         (   \beta_{3}^{ \z_{3}  }  )^2     \beta_1^{(1)} {},&\EQN{b7z8}
  \end{flalign} \vspace{\myl}

      \begin{flalign}
 & \beta_{7}^{ \z_{10}  } =  
        \frac{15}{4}\,    \beta_1^{(1)}  \beta_{6}^{ \z_{9}  } {},&\EQN{b7z10}
  \end{flalign} \vspace{\myl}

      \begin{flalign}
 & \beta_{7}^{ \z_{4} \z_{3}  } =  
        \frac{5}{2}\,    \beta_1^{(1)}  \beta_{6}^{ \z_{3}^2  } +
\frac{9}{5}\,    \beta_2^{(1)}  \beta_{5}^{ \z_{3}^2  }   -
\frac{1}{8}\,  \beta_{3}^{ \z_{3}  }  \beta_{4}^{ \z_{3}  } {},&\EQN{b7z4z3}
  \end{flalign} \vspace{\myl}

      \begin{flalign}
 & \beta_{7}^{ \z_{4} \z_{5}  } =  
           -
\frac{25}{4}\,    \beta_1^{(1)}  \beta_{6}^{ \ovl{\z}_{5,3}  } +
\frac{5}{4}\,    \beta_1^{(1)}  \beta_{6}^{ \z_{3} \z_{5}  }   -
\frac{1}{2}\,  \beta_{3}^{ \z_{3}  }  \beta_{4}^{ \z_{5}  } {},&\EQN{b7z4z5}
  \end{flalign} \vspace{\myl}

      \begin{flalign}
 & \beta_{7}^{ \z_{4} \z_{7}  } =  0 {},&\EQN{b7z4z7}
  \end{flalign} \vspace{\myl}

      \begin{flalign}
 & \beta_{7}^{ \z_{6} \z_{3}  } =  
        \frac{25}{12}\,    \beta_1^{(1)}  \beta_{6}^{ \z_{3} \z_{5}  } +
\frac{5}{8}\,  \beta_{3}^{ \z_{3}  }  \beta_{4}^{ \z_{5}  } {},&\EQN{b7z6z3}
  \end{flalign} \vspace{\myl}

      \begin{flalign}
 & \beta_{7}^{ \z_{6} \z_{5}  } =  0 {},&\EQN{b7z6z5}
  \end{flalign} \vspace{\myl}

      \begin{flalign}
 & \beta_{7}^{ \z_{8} \z_{3}  } =  0 {},&\EQN{b7z8z3}
  \end{flalign} \vspace{\myl}

      \begin{flalign}
 & \beta_{7}^{ \z_{4} \z_{3}^2  } =  
        \frac{15}{4}\,    \beta_1^{(1)}  \beta_{6}^{ \z_{3}^3  } {}
{}.
&\EQN{b7z4z3z3}
  \end{flalign} 

%\vspace{\myl}

%We would  like to stress that our predictions \re{g41}--\re{b79}
%are only  valid provided 

\subsection{Tests at 7 loops}

With eqs. \re{a4z4}--\re{b7z4z3z3} we have been able to reproduce
successfully all $\pi$-dependent constants appearing in the
$\beta$-function and anomalous dimensions $\gamma_m$ and $\gamma_2$ of
the $O(n)$ $\varphi^4$ model which all are known at 7 loops from
\cite{Schnetz:2016fhy}. In addition, we have checked that the
$\pi$-dependent contributions to the terms of order $n_f^6\alpha_s^7$
in  the QCD $\beta$-function as well as to the terms of order
$n_f^6\alpha_s^7$ and of order $n_f^5\alpha_s^7$ contributing to the
quark mass AD (all computed in
\cite{Gracey:1996he,Ciuchini:1999cv,Ciuchini:1999wy}) within the framework of large $N_f$
\cite{ Vasiliev:1981yc,Vasiliev:1981dg,Vasiliev:1982dc,Broadhurst:1996ur,Gracey:2018ame}
approach  are in full  agreement
with the constraints listed above.

Numerous successful tests  at  4,5 and 6 loops 
have been presented in \cite{Baikov:2018wgs}.

\section{\EQN{pi12}Hatted representation for   7-loop  p-integrals  and  the $\pi^{12}$ subtlety}

Motivated by the success of our derivation of the hatted representation for the 6-loop case
we have  decided to look on the next, 7-loop level. Within  our approach this requires the knowledge of
the $\ep$-expansion of the 4-loop master integrals presented in \cite{Lee:2011jt}
up to the transcendental  weight  13.
In principle, the methods employed by R.~Lee and A.~and V.~Smirnovs are
powerful enough to find such an  expansion. One of the authors of
\cite{Lee:2011jt}  has provided us with  $\ep$-expansions for all
4-loop master p-integrals up to  and including  \mbox{weight 13}.

In fact, we have (well, almost) succeeded in  constructing  the hatted
representation for the subset  ${\cal P}_4/\ep^3$  of ${\cal P}_7$. Our results are
presented   below\footnote{Note that  the hatted representation of single odd zetas displayed in eqs.
(\re{hz3t13},\re{hz5t13},\re{hz7t13},\re{hz9t13},\re{hz11t13} and \re{hz13t13}
is in agreement with the recent findings of   \cite{Kotikov:2019bqo}.}.

\begin{flalign}
  \Unb{\hat{\zeta}_3}{7} \defas \Unb{\hat{\zeta}_3}{6}
  +\frac{1705\, \ep^9}{2}\, \zeta_{12}
{},
  &&
  {}
\EQN{hz3t13}
\end{flalign}

\vspace{-5mm}
\begin{flalign}
  \Unb{\hat{\zeta}_5}{7}  \defas \Unb{\hat{\zeta}_5}{6}
  - \frac{2805\, \ep^7}{4}\, \zeta_{12} 
{},
  &&
\EQN{hz5t13}
\end{flalign}
%- 2805/4*ep**7*z12
\vspace{-5mm}
\begin{flalign}
  \Unb{\hat{\zeta}_7}{7}  \defas \Unb{\hat{\zeta}_7}{6}
   + 231  \ep^5  \zeta_{12} 
{},
  &&
\EQN{hz7t13}
\end{flalign}
\vspace{-5mm}
\begin{flalign} \Unb{\hat{ \z}_{5,3} }{7}    \defas 
\Unb{
 \hat{ \z}_{5,3}  
}{6}
+ ?  \zeta_{12} 
       - \frac{105\ep^5}{2}  \zeta_{5}   \zeta_{8} 
      {}, &&\EQN{hzz53t13}
\end{flalign} \vspace{\myl}

\vspace{1mm}
\begin{flalign}
  \Unb{\hat{\zeta}_9}{7}  \defas \Unb{\hat{\zeta}_9}{6}
  - \frac{165\ep^3}{4}  \zeta_{12} 
{},
  &&
\EQN{hz9t13}
\end{flalign}
\vspace{-3mm}
\bfl
\Unb{\hat{\z}_{7,3}}{7}\defas \Unb{\z_{7,3}}{6}
- ? \z_{12} +  \ep^3\, (\frac{105}{2}\, \z_5  \z_8  + 35 \,\z_6  \z_7 )
{},
 &&
\EQN{hz73t13}
\end{flalign}

\vspace{-5mm}
\begin{flalign}
  \Unb{\hat{\zeta}_{11}}{7}  \defas \Unb{\hat{\zeta}_{11}}{6}
 + \frac{11\ep}{2}   \zeta_{12} 
{},
  &&
\EQN{hz11t13}
\end{flalign}
\vspace*{-10mm}

\begin{flalign}
  \Unb{\hat{\zeta}_{5,3,3}}{7}  \defas \Unb{\hat{\zeta}_{5,3,3}}{6}
 +? \zeta_{12} 
       + \frac{3\ep}{2}   \zeta_{4}   \z_{5,3}  
       - \frac{105\ep^2}{16}  \zeta_{5}   \zeta_{8} 
      ,&&\EQN{hzz533t13}
    \end{flalign}
%\vspace{3mm}

\begin{flalign}\Unb{ \hat{ \z}_{9,3}}{7}     \defas 
  \fbox
  {$\z_{9,3}$  }
  +? \zeta_{12} 
       - \frac{75\ep}{2}   \zeta_{6}   \zeta_{7} 
       - 21 \ep   \zeta_{5}   \zeta_{8} 
       - \frac{81\ep}{2}   \zeta_{4}   \zeta_{9} 
      ,&&\EQN{hzz93t13}
 \end{flalign} \vspace{\myl}
 \vspace{3mm}

\begin{flalign} \Unb{\hat{\zeta}_{13}}{7}    \defas 
         \fbox{$\zeta_{13}$} 
      ,&&\EQN{hz13t13}
 \end{flalign} \vspace{\myl}
 \vspace{3mm}

\begin{flalign}\Unb{ \hat{ \z}_{5,5,3}}{7}     \defas 
       \fbox{$  \z_{5,5,3}  
       + \frac{145}{12}  \zeta_{5}   \zeta_{8} 
       + 25  \zeta_{4}   \zeta_{9} 
       + \frac{275}{2}  \zeta_{2}   \zeta_{11}
       $}
           ,&&\EQN{hzz553t13}
 \end{flalign} \vspace{\myl}
 \vspace{3mm}

\begin{flalign} \Unb{\hat{ \z}_{7,3,3}}{7}     \defas 
         \fbox{$\z_{7,3,3}  
       - 4  \zeta_{6}   \zeta_{7} 
       + \frac{29}{2}  \zeta_{5}   \zeta_{8} 
       + 28  \zeta_{4}   \zeta_{9} 
       + \frac{407}{2}  \zeta_{2}   \zeta_{11}
       $}
      ,&&\EQN{hzz733t13}
 \end{flalign} \vspace{\myl}
\vspace{3mm}

\begin{flalign} \Unb{\hat{ \z}_{6,4,1,1}}{7}     \defas 
       \fbox{$  \z_{6,4,1,1}  
       + ?  \zeta_{12} 
       - \frac{3}{2}  \zeta_{4}   \z_{5,3}  
       + \frac{1}{2}  \zeta_{3}   \zeta_{4}   \zeta_{5} 
       + \frac{9}{4}   
    \z_{3}^2  \zeta_{6} 
       - 3  \zeta_{2}   \z_{7,3}  
       - \frac{7}{2}  \zeta_{2}
       \z_{5}^2
       - 10  \zeta_{2}   \zeta_{3}   \zeta_{7} 
          $}
       \nnb  &&
        \end{flalign} \vspace{\myl}      
\vspace{1.5mm}
\vspace{-2mm}

\begin{flalign}
  &&
            + \ep\left( \frac{5665 }{32}   \zeta_{6}   \zeta_{7} 
       + \frac{203 }{2}   \zeta_{5}   \zeta_{8} 
       + \frac{1799 }{12}   \zeta_{4}   \zeta_{9} 
       - \frac{799 }{16}   \zeta_{3}   \zeta_{10} 
       + \frac{1 }{2}    
       \z_{3}^3  \zeta_{4}
        \right)
                        {}.
%=
       \EQN{hzz6411t13}
       {}
\end{flalign}

%- 105/16*ep**2*z5*z8 + ep*z4*(5/2*phi - 5/2*z3*z5)

The meaning of the question mark in front of $\zeta_{12}$ in 
eqs. \re{hzz53t13},   \re{hzz533t13}, \re{hzz93t13} and \re{hzz6411t13}
for hatted form of multiple zeta objects is as follows. Every integral
from the set ${\cal P}_{4}/\ep^3$ can either include at least one (or
more) multiple zeta values from the collection
$
\oz_{5,3}, \oz_{7,3}, \oz_{9,3},\oz_{5,3,3},$
$\oz_{5,5,3}, \oz_{7,3,3}$
and $\oz_{6,4,1,1}$ 
or not. \ice{ (Please, keep in mind,
that for a given p-integral we consider only the pole and constant
parts in its $\ep$ expansion.)} Thus, the whole set
${\cal P}_{4}/\ep^3$ can be represented as a union of two
(non-intersecting!) subsets, namely, a simple one,
${\cal S}_{4}/\ep^3$, (that is without any dependence on multiple zeta values)
and the rest ${\cal N }_{4}/\ep^3$.

The fact is that the hatted representation does exists for all
p-integrals ${\cal S}_{4}/\ep^3$, while there is no way to replace the
question marks in eqs above by some coefficients in order  to meet
eq.~\re{hat:master} for the p-integrals from ${\cal N }_{4}/\ep^3$. On
the other hand, if we formally set to zero all terms proportional to $\z_{12}$
in eqs. (\ref{hz3t13}-\ref{hzz6411t13}), then
eq.~\re{hat:master} will be valid for the whole set
${\cal P}_{4}/\ep^3$ ``modulo'' terms proportional to $\zeta_{12}$.

It is quite remarkable that the distinguished role of $\z_{12}$ has
been already established in \cite{Schnetz:2016fhy} as a result of direct analytical
calculations of quite complicated {\em convergent} 7-loop p-integrals.
\ice{(which essentially  coincide with 8-loop Feynman periods).}

Thus, we observe  a nontrivial interplay between higher terms in the
$\ep$-expansion of 4-loop p-integrals and 7-loop {\em finite}  p-integrals.

Certainly, the subset of the 7-loop p-integrals which has led to
eqs.~(\ref{hz3t13}-\ref{hzz6411t13}) is rather limited and our conclusions
about $\pi$-structure of ${\cal P}_7$ are not final. In principle, we
can outline 3 possible scenarios.  \vspace{2mm}

\noindent
{\bf Scenario 1} (\normalsize pessimistic). There is no
hatted  representation  for the set ${\cal P}_7 $.

\vspace{2mm}

\noindent
{\bf Scenario 2} (conservative). The  master p-integrals
from the difference ${\cal P}_7 \ \backslash \  {\cal P}_4/\ep^3$
can be  presented in the hatted form modulo (explicitly) $\pi$-proportional
terms with  weight  more  or equal 12.
\vspace{2mm} 

\noindent
{\bf Scenario 3} (optimistic).
The master p-integrals
from the difference ${\cal P}_7 \  \backslash  \ {\cal P}_4/\ep^3$
can be  presented in the hatted form modulo $\zeta_{12}$.

\section{\EQN{8l:results} $\pi$-dependence of 8-loop $\beta$-functions and AD}

In this section we assume the conservative  Scenario 2 and  extend (following generic
prescriptions elaborated in \cite{{Baikov:2018wgs}}) the
predictions from Section 4 by one more  loop for $\pi$-dependent  
terms with the transcendental weight  not exceeding 11.  The results read:

      \begin{flalign}
 & \gamma_{8}^{ \z_{4}  } =  
           -
\frac{3}{14}\,    \gamma_1^{(1)}  \beta_{7}^{ \z_{3}  }   -
\frac{1}{2}\,    \gamma_2^{(1)}  \beta_{6}^{ \z_{3}  }   -
\frac{9}{10}\,    \gamma_3^{(1)}  \beta_{5}^{ \z_{3}  } +
\frac{3}{2}\,  \gamma_{3}^{ \z_{3}  }    \beta_5^{(1)}   -
\frac{3}{2}\,    \gamma_4^{(1)}  \beta_{4}^{ \z_{3}  } 
\brk
         +
\frac{3}{2}\,  \gamma_{4}^{ \z_{3}  }    \beta_4^{(1)}   -
\frac{5}{2}\,    \gamma_5^{(1)}  \beta_{3}^{ \z_{3}  } +
\frac{3}{2}\,  \gamma_{5}^{ \z_{3}  }    \beta_3^{(1)} +
\frac{3}{2}\,  \gamma_{6}^{ \z_{3}  }    \beta_2^{(1)} +
\frac{3}{2}\,  \gamma_{7}^{ \z_{3}  }    \beta_1^{(1)} {},&\EQN{a8z4}
  \end{flalign} \vspace{\myl}

      \begin{flalign}
 & \gamma_{8}^{ \z_{6}  } =  
           -
\frac{5}{14}\,    \gamma_1^{(1)}  \beta_{7}^{ \z_{5}  }   -
\frac{5}{6}\,    \gamma_2^{(1)}  \beta_{6}^{ \z_{5}  }   -
\frac{3}{2}\,    \gamma_3^{(1)}  \beta_{5}^{ \z_{5}  }   -
\frac{5}{2}\,    \gamma_4^{(1)}  \beta_{4}^{ \z_{5}  } +
\frac{5}{2}\,    \beta_4^{(1)}  \gamma_{4}^{ \z_{5}  } 
         +
\frac{5}{2}\,    \beta_3^{(1)}  \gamma_{5}^{ \z_{5}  } 
\brk
+
\frac{5}{2}\,    \beta_2^{(1)}  \gamma_{6}^{ \z_{5}  } +
\frac{15}{14}\,  (     \beta_2^{(1)}  )^2   \beta_{3}^{ \z_{3}  }    \gamma_1^{(1)} +
\frac{5}{2}\,    \beta_1^{(1)}  \gamma_{7}^{ \z_{5}  }
\brk
 +
\frac{15}{7}\,  
           \beta_1^{(1)}    \beta_3^{(1)}  \beta_{3}^{ \z_{3}  }    \gamma_1^{(1)} +
\frac{15}{7}\,    \beta_1^{(1)}    \beta_2^{(1)}  \beta_{4}^{ \z_{3}  }    \gamma_1^{(1)} +
5    \beta_1^{(1)}    \beta_2^{(1)}  \beta_{3}^{ \z_{3}  }    \gamma_2^{(1)}   -
\frac{15}{2}\,    \beta_1^{(1)}  (     \beta_2^{(1)}  )^2   \gamma_{3}^{ \z_{3}  } 
\brk
         +
\frac{15}{14}\,  (     \beta_1^{(1)}  )^2   \beta_{5}^{ \z_{3}  }    \gamma_1^{(1)} +
\frac{5}{2}\,  (     \beta_1^{(1)}  )^2   \beta_{4}^{ \z_{3}  }    \gamma_2^{(1)} +
\frac{9}{2}\,  (     \beta_1^{(1)}  )^2   \beta_{3}^{ \z_{3}  }    \gamma_3^{(1)}   -
\frac{15}{2}\,  (     \beta_1^{(1)}  )^2   
           \beta_3^{(1)}  \gamma_{3}^{ \z_{3}  } 
\brk   -
\frac{15}{2}\,  (     \beta_1^{(1)}  )^2     \beta_2^{(1)}  \gamma_{4}^{ \z_{3}  }   -
\frac{5}{2}\,  (     \beta_1^{(1)}  )^3   \gamma_{5}^{ \z_{3}  } {},&\EQN{a8z6}
  \end{flalign} \vspace{\myl}

      \begin{flalign}
 & \gamma_{8}^{ \z_{8}  } =  
           -
\frac{1}{2}\,    \gamma_1^{(1)}  \beta_{7}^{ \z_{7}  }   -
\frac{7}{6}\,    \gamma_2^{(1)}  \beta_{6}^{ \z_{7}  }   -
\frac{21}{10}\,    \gamma_3^{(1)}  \beta_{5}^{ \z_{7}  } +
\beta_{3}^{ \z_{3}  }  \beta_{4}^{ \z_{3}  }    \gamma_1^{(1)} +
\frac{35}{24}\,  
         (   \beta_{3}^{ \z_{3}  }  )^2     \gamma_2^{(1)} 
\brk
+
\frac{7}{2}\,    \beta_3^{(1)}  \gamma_{5}^{ \z_{7}  } +
\frac{7}{2}\,    \beta_2^{(1)}  \gamma_{6}^{ \z_{7}  }   -
\frac{3}{4}\,    \beta_2^{(1)}    \gamma_1^{(1)}  \beta_{5}^{ \z_{3}^2  }   -
\frac{7}{4}\,    \beta_2^{(1)}  \beta_{3}^{ \z_{3}  }  
         \gamma_{3}^{ \z_{3}  } 
\brk
+
\frac{7}{2}\,    \beta_1^{(1)}  \gamma_{7}^{ \z_{7}  }   -
\frac{3}{4}\,    \beta_1^{(1)}    \gamma_1^{(1)}  \beta_{6}^{ \z_{3}^2  }   -
\frac{7}{4}\,    \beta_1^{(1)}    \gamma_2^{(1)}  \beta_{5}^{ \z_{3}^2  }   -
\frac{7}{4}\,    \beta_1^{(1)}  \beta_{3}^{ \z_{3}  }  
         \gamma_{4}^{ \z_{3}  }
\brk
 +
\frac{21}{4}\,    \beta_1^{(1)}    \beta_2^{(1)}  \gamma_{5}^{ \z_{3}^2  } +
\frac{15}{2}\,    \beta_1^{(1)}    \beta_2^{(1)}    \gamma_1^{(1)}  \beta_{4}^{ \z_{5}  } +
\frac{21}{8}\,  (     \beta_1^{(1)}  )^2   \gamma_{6}^{ \z_{3}^2  } +
\frac{15}{4}\,  
         (     \beta_1^{(1)}  )^2     \gamma_1^{(1)}  \beta_{5}^{ \z_{5}  } 
\brk
+
\frac{35}{4}\,  (     \beta_1^{(1)}  )^2     \gamma_2^{(1)}  \beta_{4}^{ \z_{5}  }   -
\frac{105}{4}\,  (     \beta_1^{(1)}  )^2     \beta_2^{(1)}  \gamma_{4}^{ \z_{5}  }   -
\frac{35}{4}\,  (     \beta_1^{(1)}  )^3   \gamma_{5}^{ \z_{5}  } 
\brk    
       -
\frac{15}{2}\,  (     \beta_1^{(1)}  )^4   \beta_{3}^{ \z_{3}  }    \gamma_1^{(1)} +
\frac{21}{2}\,  (     \beta_1^{(1)}  )^5   \gamma_{3}^{ \z_{3}  } {},&\EQN{a8z8}
  \end{flalign} \vspace{\myl}

      \begin{flalign}
 & \gamma_{8}^{ \z_{10}  } =  
           -
\frac{9}{14}\,    \gamma_1^{(1)}  \beta_{7}^{ \z_{9}  }   -
\frac{3}{2}\,    \gamma_2^{(1)}  \beta_{6}^{ \z_{9}  } +
\frac{11}{7}\,  \beta_{3}^{ \z_{3}  }    \gamma_1^{(1)}  \beta_{4}^{ \z_{5}  } +
\frac{9}{2}\,    \beta_2^{(1)}  \gamma_{6}^{ \z_{9}  } +
\frac{9}{2}\,    \beta_1^{(1)} 
          \gamma_{7}^{ \z_{9}  } +
\frac{415}{188}\,    \beta_1^{(1)}    \gamma_1^{(1)}  \beta_{6}^{ \ovl{\z}_{5,3}  }   
\brk
-
\frac{33}{28}\,    \beta_1^{(1)}    \gamma_1^{(1)}  \beta_{6}^{ \z_{3} \z_{5}  }   -
\frac{11}{4}\,    \beta_1^{(1)}  \beta_{3}^{ \z_{3}  }  \gamma_{4}^{ \z_{5}  }   -
\frac{2905}{376}\,  (     \beta_1^{(1)}  )^2   \gamma_{6}^{ \ovl{\z}_{5,3}  }
\brk
 +
\frac{33}{8}\,  (     \beta_1^{(1)}  )^2   \gamma_{6}^{ \z_{3} \z_{5}  } +
9  (     \beta_1^{(1)}  )^2     \gamma_1^{(1)}  \beta_{5}^{ \z_{7}  }   -
21  (     \beta_1^{(1)}  )^3   \gamma_{5}^{ \z_{7}  } {},&\EQN{a8z10}
  \end{flalign} \vspace{\myl}

      \begin{flalign}
 & \gamma_{8}^{ \z_{4} \z_{3}  } =  
           -
\frac{3}{7}\,    \gamma_1^{(1)}  \beta_{7}^{ \z_{3}^2  }   -
\gamma_2^{(1)}  \beta_{6}^{ \z_{3}^2  }   -
\frac{9}{5}\,    \gamma_3^{(1)}  \beta_{5}^{ \z_{3}^2  } +
\frac{3}{5}\,  \gamma_{3}^{ \z_{3}  }  \beta_{5}^{ \z_{3}  } +
3    \beta_3^{(1)}  
         \gamma_{5}^{ \z_{3}^2  }
\brk
   -
\gamma_{5}^{ \z_{3}  }  \beta_{3}^{ \z_{3}  } +
3    \beta_2^{(1)}  \gamma_{6}^{ \z_{3}^2  } +
3    \beta_1^{(1)}  \gamma_{7}^{ \z_{3}^2  } {},&\EQN{a8z4z3}
  \end{flalign} \vspace{\myl}

      \begin{flalign}
 & \gamma_{8}^{ \z_{4} \z_{5}  } =  
        \frac{15}{14}\,    \gamma_1^{(1)}  \beta_{7}^{ \ovl{\z}_{5,3}  }   -
\frac{3}{14}\,    \gamma_1^{(1)}  \beta_{7}^{ \z_{3} \z_{5}  } +
\frac{5}{2}\,    \gamma_2^{(1)}  \beta_{6}^{ \ovl{\z}_{5,3}  }   -
\frac{1}{2}\,    \gamma_2^{(1)}  \beta_{6}^{ \z_{3} \z_{5}  } +
\frac{3}{2}\,  
         \gamma_{3}^{ \z_{3}  }  \beta_{5}^{ \z_{5}  }  
\brk
 -
\frac{3}{2}\,  \beta_{4}^{ \z_{3}  }  \gamma_{4}^{ \z_{5}  } +
\frac{3}{2}\,  \gamma_{4}^{ \z_{3}  }  \beta_{4}^{ \z_{5}  }   -
\frac{5}{2}\,  \beta_{3}^{ \z_{3}  }  \gamma_{5}^{ \z_{5}  }   -
\frac{15}{2}\,    \beta_2^{(1)}  
         \gamma_{6}^{ \ovl{\z}_{5,3}  }
\brk
 +
\frac{3}{2}\,    \beta_2^{(1)}  \gamma_{6}^{ \z_{3} \z_{5}  }   -
\frac{15}{2}\,    \beta_1^{(1)}  \gamma_{7}^{ \ovl{\z}_{5,3}  } +
\frac{3}{2}\,    \beta_1^{(1)}  \gamma_{7}^{ \z_{3} \z_{5}  } {},&\EQN{a8z4z5}
  \end{flalign} \vspace{\myl}

      \begin{flalign}
 & \gamma_{8}^{ \z_{4} \z_{7}  } =  
        3    \gamma_1^{(1)}  \beta_{7}^{ \ovl{\z}_{7,3}  }   -
\frac{3}{14}\,    \gamma_1^{(1)}  \beta_{7}^{ \z_{3} \z_{7}  } +
\frac{3}{2}\,  \gamma_{3}^{ \z_{3}  }  \beta_{5}^{ \z_{7}  }   -
\frac{5}{2}\,  \beta_{3}^{ \z_{3}  }  \gamma_{5}^{ \z_{7}  }   -
21    \beta_1^{(1)}  
         \gamma_{7}^{ \ovl{\z}_{7,3}  } +
\frac{3}{2}\,    \beta_1^{(1)}  \gamma_{7}^{ \z_{3} \z_{7}  } {},&\EQN{a8z4z7}
  \end{flalign} \vspace{\myl}

\ice{   
      \begin{flalign}
 & \gamma_{8}^{ \z_{4} \z_{9}  } =  0 {},&\EQN{a8z4z9}
  \end{flalign} \vspace{\myl}
}

      \begin{flalign}
 & \gamma_{8}^{ \z_{6} \z_{3}  } =  
           -
\frac{5}{14}\,    \gamma_1^{(1)}  \beta_{7}^{ \z_{3} \z_{5}  }   -
\frac{5}{6}\,    \gamma_2^{(1)}  \beta_{6}^{ \z_{3} \z_{5}  }   -
\frac{3}{2}\,  \gamma_{3}^{ \z_{3}  }  \beta_{5}^{ \z_{5}  } +
\frac{5}{2}\,  \beta_{4}^{ \z_{3}  }  \gamma_{4}^{ \z_{5}  }   -
\frac{5}{2}\,  
         \gamma_{4}^{ \z_{3}  }  \beta_{4}^{ \z_{5}  } 
\brk
+
\frac{5}{2}\,  \beta_{3}^{ \z_{3}  }  \gamma_{5}^{ \z_{5}  } +
\frac{5}{2}\,    \beta_2^{(1)}  \gamma_{6}^{ \z_{3} \z_{5}  } +
\frac{5}{2}\,    \beta_1^{(1)}  \gamma_{7}^{ \z_{3} \z_{5}  } +
\frac{15}{7}\,    \beta_1^{(1)}  
         (   \beta_{3}^{ \z_{3}  }  )^2     \gamma_1^{(1)} 
\brk
+
\frac{15}{7}\,  (     \beta_1^{(1)}  )^2     \gamma_1^{(1)}  \beta_{5}^{ \z_{3}^2  }   -
3  (     \beta_1^{(1)}  )^2   \beta_{3}^{ \z_{3}  }  \gamma_{3}^{ \z_{3}  }   -
5  (     \beta_1^{(1)}  )^3   \gamma_{5}^{ \z_{3}^2  } {},&\EQN{a8z6z3}
  \end{flalign} \vspace{\myl}

      \begin{flalign}
 & \gamma_{8}^{ \z_{6} \z_{5}  } =  
        \frac{15}{7}\,    \gamma_1^{(1)}  \beta_{7}^{ \ovl{\z}_{7,3}  }   -
\frac{5}{7}\,    \gamma_1^{(1)}  \beta_{7}^{ \z_{5}^2  }   -
15    \beta_1^{(1)}  \gamma_{7}^{ \ovl{\z}_{7,3}  } +
5    \beta_1^{(1)}  \gamma_{7}^{ \z_{5}^2  } {},&\EQN{a8z6z5}
  \end{flalign} \vspace{\myl}

 \ice{
      \begin{flalign}
 & \gamma_{8}^{ \z_{6} \z_{7}  } =  0 {},&\EQN{a8z6z7}
  \end{flalign} \vspace{\myl}
   }

      \begin{flalign}
 & \gamma_{8}^{ \z_{8} \z_{3}  } =  
           -
\frac{1}{2}\,    \gamma_1^{(1)}  \beta_{7}^{ \z_{3} \z_{7}  }   -
\frac{21}{10}\,  \gamma_{3}^{ \z_{3}  }  \beta_{5}^{ \z_{7}  } +
\frac{7}{2}\,  \beta_{3}^{ \z_{3}  }  \gamma_{5}^{ \z_{7}  } +
\frac{7}{2}\,    \beta_1^{(1)}  \gamma_{7}^{ \z_{3} \z_{7}  }   -
\frac{9}{4}\, 
            \beta_1^{(1)}    \gamma_1^{(1)}  \beta_{6}^{ \z_{3}^3  } +
\frac{63}{8}\,  (     \beta_1^{(1)}  )^2   \gamma_{6}^{ \z_{3}^3  } {},&\EQN{a8z8z3}
  \end{flalign} \vspace{\myl}

\ice{
      \begin{flalign}
 & \gamma_{8}^{ \z_{8} \z_{5}  } =  0 {},&\EQN{a8z8z5}
  \end{flalign} \vspace{\myl}

      \begin{flalign}
 & \gamma_{8}^{ \z_{10} \z_{3}  } =  0 {},&\EQN{a8z10z3}
  \end{flalign} \vspace{\myl}

      \begin{flalign}
 & \gamma_{8}^{ \z_{4} \z_{3}^2  } =  
           -
\frac{9}{14}\,    \gamma_1^{(1)}  \beta_{7}^{ \z_{3}^3  }   -
\frac{3}{2}\,    \gamma_2^{(1)}  \beta_{6}^{ \z_{3}^3  }   -
\frac{3}{10}\,  \gamma_{3}^{ \z_{3}  }  \beta_{5}^{ \z_{3}^2  } +
\frac{1}{2}\,  \beta_{3}^{ \z_{3}  }  
         \gamma_{5}^{ \z_{3}^2  } +
\frac{9}{2}\,    \beta_2^{(1)}  \gamma_{6}^{ \z_{3}^3  } +
\frac{9}{2}\,    \beta_1^{(1)}  \gamma_{7}^{ \z_{3}^3  } {},&\EQN{a8z4z3z3}
  \end{flalign} \vspace{\myl}

      \begin{flalign}
 & \gamma_{8}^{ \z_{4} \z_{3} \z_{5}  } =  
        \frac{15}{14}\,    \gamma_1^{(1)}  \beta_{7}^{ \z_{3} \ovl{\z}_{5,3}  }   -
\frac{3}{7}\,    \gamma_1^{(1)}  \beta_{7}^{ \z_{3}^2 \z_{5}  }   -
\frac{15}{2}\,    \beta_1^{(1)}  \gamma_{7}^{ \z_{3} \ovl{\z}_{5,3}  } +
3    \beta_1^{(1)}  \gamma_{7}^{ \z_{3}^2 \z_{5}  } {},&\EQN{a8z4z3z5}
  \end{flalign} \vspace{\myl}

      \begin{flalign}
 & \gamma_{8}^{ \z_{4} \ovl{\z}_{5,3}  } =  
           -
\frac{3}{14}\,    \gamma_1^{(1)}  \beta_{7}^{ \ovl{\z}_{5,3,3}  }   -
\frac{3}{14}\,    \gamma_1^{(1)}  \beta_{7}^{ \z_{3} \ovl{\z}_{5,3}  } +
\frac{3}{2}\,    \beta_1^{(1)}  \gamma_{7}^{ \ovl{\z}_{5,3,3}  } +
\frac{3}{2}\,    \beta_1^{(1)}  \gamma_{7}^{ \z_{3} \ovl{\z}_{5,3}  } 
         {},&\EQN{a8z4zz53}
  \end{flalign} \vspace{\myl}

      \begin{flalign}
 & \gamma_{8}^{ \z_{6} \z_{3}^2  } =  
           -
\frac{5}{14}\,    \gamma_1^{(1)}  \beta_{7}^{ \z_{3}^2 \z_{5}  } +
\frac{5}{2}\,    \beta_1^{(1)}  \gamma_{7}^{ \z_{3}^2 \z_{5}  } {},&\EQN{a8z6z3z3}
  \end{flalign} \vspace{\myl}

      \begin{flalign}
 & \gamma_{8}^{ \z_{4} \z_{3}^3  } =  0 {},&\EQN{a8z4z3z3z3}
  \end{flalign} \vspace{\myl}
   
}

      \begin{flalign}
 & \beta_{8}^{ \z_{4}  } =  
        \frac{9}{7}\,    \beta_1^{(1)}  \beta_{7}^{ \z_{3}  } +
\beta_2^{(1)}  \beta_{6}^{ \z_{3}  } +
\frac{3}{5}\,    \beta_3^{(1)}  \beta_{5}^{ \z_{3}  }   -
\beta_{3}^{ \z_{3}  }    \beta_5^{(1)} {},&\EQN{b8z4}
  \end{flalign} \vspace{\myl}

      \begin{flalign}
 & \beta_{8}^{ \z_{6}  } =  
        \frac{15}{7}\,    \beta_1^{(1)}  \beta_{7}^{ \z_{5}  } +
\frac{5}{3}\,    \beta_2^{(1)}  \beta_{6}^{ \z_{5}  }   -
\frac{20}{7}\,    \beta_2^{(1)}  (     \beta_1^{(1)}  )^2   \beta_{4}^{ \z_{3}  }   -
\frac{10}{7}\,  \beta_{5}^{ \z_{3}  }  (     \beta_1^{(1)}  )^3  
\brk
+
\beta_3^{(1)}  
         \beta_{5}^{ \z_{5}  }  
 -
\frac{10}{7}\,  \beta_{3}^{ \z_{3}  }  (     \beta_2^{(1)}  )^2     \beta_1^{(1)}   -
\frac{6}{7}\,  \beta_{3}^{ \z_{3}  }    \beta_3^{(1)}  (     \beta_1^{(1)}  )^2  {},&\EQN{b8z6}
  \end{flalign} \vspace{\myl}

      \begin{flalign}
 & \beta_{8}^{ \z_{8}  } =  
        3    \beta_1^{(1)}  \beta_{7}^{ \z_{7}  } +
\frac{15}{8}\,  (     \beta_1^{(1)}  )^2   \beta_{6}^{ \z_{3}^2  }   -
5  (     \beta_1^{(1)}  )^3   \beta_{5}^{ \z_{5}  } +
\frac{7}{3}\,    \beta_2^{(1)}  \beta_{6}^{ \z_{7}  } +
\frac{11}{4}\,    \beta_2^{(1)}    \beta_1^{(1)}  
         \beta_{5}^{ \z_{3}^2  } 
\brk
  -
10    \beta_2^{(1)}  (     \beta_1^{(1)}  )^2   \beta_{4}^{ \z_{5}  } +
\frac{7}{5}\,    \beta_3^{(1)}  \beta_{5}^{ \z_{7}  }   -
\frac{3}{4}\,  \beta_{3}^{ \z_{3}  }    \beta_1^{(1)}  \beta_{4}^{ \z_{3}  } +
3  \beta_{3}^{ \z_{3}  }  (     \beta_1^{(1)}  )^5  
           -
\frac{7}{24}\,  (   \beta_{3}^{ \z_{3}  }  )^2     \beta_2^{(1)} {},&\EQN{b8z8}
  \end{flalign} \vspace{\myl}

      \begin{flalign}
 & \beta_{8}^{ \z_{10}  } =  
        \frac{27}{7}\,    \beta_1^{(1)}  \beta_{7}^{ \z_{9}  }   -
\frac{2075}{376}\,  (     \beta_1^{(1)}  )^2   \beta_{6}^{ \ovl{\z}_{5,3}  } +
\frac{165}{56}\,  (     \beta_1^{(1)}  )^2   \beta_{6}^{ \z_{3} \z_{5}  }   
\brk
-
12  (     \beta_1^{(1)}  )^3   \beta_{5}^{ \z_{7}  } 
         +
3    \beta_2^{(1)}  \beta_{6}^{ \z_{9}  }   -
\frac{33}{28}\,  \beta_{3}^{ \z_{3}  }    \beta_1^{(1)}  \beta_{4}^{ \z_{5}  } {},&\EQN{b8z10}
  \end{flalign} \vspace{\myl}

      \begin{flalign}
 & \beta_{8}^{ \z_{4} \z_{3}  } =  
        \frac{18}{7}\,    \beta_1^{(1)}  \beta_{7}^{ \z_{3}^2  } +
2    \beta_2^{(1)}  \beta_{6}^{ \z_{3}^2  } +
\frac{6}{5}\,    \beta_3^{(1)}  \beta_{5}^{ \z_{3}^2  }   -
\frac{2}{5}\,  \beta_{3}^{ \z_{3}  }  \beta_{5}^{ \z_{3}  } {},&\EQN{b8z4z3}
  \end{flalign} \vspace{\myl}

      \begin{flalign}
 & \beta_{8}^{ \z_{4} \z_{5}  } =  
           -
\frac{45}{7}\,    \beta_1^{(1)}  \beta_{7}^{ \ovl{\z}_{5,3}  } +
\frac{9}{7}\,    \beta_1^{(1)}  \beta_{7}^{ \z_{3} \z_{5}  }   -
5    \beta_2^{(1)}  \beta_{6}^{ \ovl{\z}_{5,3}  } +
\beta_2^{(1)}  \beta_{6}^{ \z_{3} \z_{5}  }   -
\beta_{3}^{ \z_{3}  }  \beta_{5}^{ \z_{5}  } {},&\EQN{b8z4z5}
  \end{flalign} \vspace{\myl}

      \begin{flalign}
 & \beta_{8}^{ \z_{4} \z_{7}  } =  
           -
18    \beta_1^{(1)}  \beta_{7}^{ \ovl{\z}_{7,3}  } +
\frac{9}{7}\,    \beta_1^{(1)}  \beta_{7}^{ \z_{3} \z_{7}  }   -
\beta_{3}^{ \z_{3}  }  \beta_{5}^{ \z_{7}  } {},&\EQN{b8z4z7}
  \end{flalign} \vspace{\myl}
   
   \ice{
   
      \begin{flalign}
 & \beta_{8}^{ \z_{4} \z_{9}  } =  0 {},&\EQN{b8z4z9}
  \end{flalign} \vspace{\myl}
  }

      \begin{flalign}
 & \beta_{8}^{ \z_{6} \z_{3}  } =  
        \frac{15}{7}\,    \beta_1^{(1)}  \beta_{7}^{ \z_{3} \z_{5}  }   -
\frac{20}{7}\,  (     \beta_1^{(1)}  )^3   \beta_{5}^{ \z_{3}^2  } +
\frac{5}{3}\,    \beta_2^{(1)}  \beta_{6}^{ \z_{3} \z_{5}  } +
\beta_{3}^{ \z_{3}  }  \beta_{5}^{ \z_{5}  }   -
\frac{6}{7}\,  
         (   \beta_{3}^{ \z_{3}  }  )^2   (     \beta_1^{(1)}  )^2  {},&\EQN{b8z6z3}
  \end{flalign} \vspace{\myl}

      \begin{flalign}
 & \beta_{8}^{ \z_{6} \z_{5}  } =  
           -
\frac{90}{7}\,    \beta_1^{(1)}  \beta_{7}^{ \ovl{\z}_{7,3}  } +
\frac{30}{7}\,    \beta_1^{(1)}  \beta_{7}^{ \z_{5}^2  } {},&\EQN{b8z6z5}
  \end{flalign} \vspace{\myl}

\ice{
   
      \begin{flalign}
 & \beta_{8}^{ \z_{6} \z_{7}  } =  0 {},&\EQN{b8z6z7}
  \end{flalign} \vspace{\myl}
}

      \begin{flalign}
 & \beta_{8}^{ \z_{8} \z_{3}  } =  
        3    \beta_1^{(1)}  \beta_{7}^{ \z_{3} \z_{7}  } +
\frac{45}{8}\,  (     \beta_1^{(1)}  )^2   \beta_{6}^{ \z_{3}^3  } +
\frac{7}{5}\,  \beta_{3}^{ \z_{3}  }  \beta_{5}^{ \z_{7}  } {},&\EQN{b8z8z3}
  \end{flalign} \vspace{\myl}

\ice{
      \begin{flalign}
 & \beta_{8}^{ \z_{8} \z_{5}  } =  0 {},&\EQN{b8z8z5}
  \end{flalign} \vspace{\myl}

      \begin{flalign}
 & \beta_{8}^{ \z_{10} \z_{3}  } =  0 {},&\EQN{b8z10z3}
  \end{flalign} \vspace{\myl}
   
}   
   
      \begin{flalign}
 & \beta_{8}^{ \z_{4} \z_{3}^2  } =  
        \frac{27}{7}\,    \beta_1^{(1)}  \beta_{7}^{ \z_{3}^3  } +
3    \beta_2^{(1)}  \beta_{6}^{ \z_{3}^3  } +
\frac{1}{5}\,  \beta_{3}^{ \z_{3}  }  \beta_{5}^{ \z_{3}^2  } {}
{}.
&\EQN{b8z4z3z3}
\end{flalign}

%\vspace{\myl}
   
 \ice{  
   
      \begin{flalign}
 & \beta_{8}^{ \z_{4} \z_{3} \z_{5}  } =  
           -
\frac{45}{7}\,    \beta_1^{(1)}  \beta_{7}^{ \z_{3} \ovl{\z}_{5,3}  } +
\frac{18}{7}\,    \beta_1^{(1)}  \beta_{7}^{ \z_{3}^2 \z_{5}  } {},&\EQN{b8z4z3z5}
  \end{flalign} \vspace{\myl}

      \begin{flalign}
 & \beta_{8}^{ \z_{4} \ovl{\z}_{5,3}  } =  
        \frac{9}{7}\,    \beta_1^{(1)}  \beta_{7}^{ \ovl{\z}_{5,3,3}  } +
\frac{9}{7}\,    \beta_1^{(1)}  \beta_{7}^{ \z_{3} \ovl{\z}_{5,3}  } {},&\EQN{b8z4zz53}
  \end{flalign} \vspace{\myl}

      \begin{flalign}
 & \beta_{8}^{ \z_{6} \z_{3}^2  } =  
        \frac{15}{7}\,    \beta_1^{(1)}  \beta_{7}^{ \z_{3}^2 \z_{5}  } {},&\EQN{b8z6z3z3}
  \end{flalign} \vspace{\myl}

      \begin{flalign}
 & \beta_{8}^{ \z_{4} \z_{3}^3  } =  0 {}.  
&\EQN{b8z4z3z3z3}
  \end{flalign} \vspace{\myl}
}

\subsection{Tests \ice{of  our results} at 8 loops}

Here we summarize  all currently  available   evidence  supporting
assumptions (that is  scenarios 2 and 3)  leading to eqs. (\ref{a8z4}--\ref{b8z4z3z3}).

First of all, we have checked that currently  \ice{all (though rather limited)}  known \ice{to us} 8-loop
results for ADs and $\beta$-functions are in full agreement to our
predictions. Namely, we have successfully  checked  the following cases.

\begin{itemize}

\item  Contributions of order    $\alpha_s^8\, N_f^7$ to the  QCD $\beta$-function \cite{Gracey:1996he}. 

\item  Contributions of orders    $\alpha_s^8\, N_f^7$ and  $\alpha_s^8\, N_f^6$ 
to the  QCD  quark mass anomalous dimension \cite{Ciuchini:1999cv,Ciuchini:1999wy}. 

\item  Contributions of order    $g^8 n^7$ and   $g^8 n^6$
 to the $\beta$-function, the field anomalous dimension and to the mass anomalous dimension
  of the  scalar $O(n)$  $\phi^4$ theory \cite{Broadhurst:1996ur}.
\end{itemize}

The above list give some support to the conservative Scenario 2.
In fact, there is  an argument in favor of even  less conservative  optimistic Scenario 3.  
Indeed, according our definitions   the  value of  any convergent ({\em and} expressible 
in terms of multiple zeta values only)
7-loop p-integral at $D=4$
should  become  completely $\pi$-free (modulo terms proportional to  $\pi^{12}$) 
if rewritten  in terms of the generators 
\beq \z_3,\z_5,\z_7, \oz_{5,3},
\z_9,  \oz_{7,3}, \z_{11}, \oz_{5,3,3}, \oz_{9,3}, \z_{13}, \oz_{5,5,3},  \oz_{7,3,3},
\oz_{6,4,1,1}
{}.
\EQN{gensL7}
\eeq

The authors of \cite{Panzer:2016snt} have published  analytic results for  a   large  collection of
finite p-integrals. At 7 loops the collection contains 369 7-loop finite  p-integrals
which depends on  single and multiple  zeta values only\footnote{ We do not count uninteresting for
our discussion cases of p-integrals depending only on single (odd)  zetas, as the latter do not
depend on  $\pi$ at all neither in original nor in  hatted forms in the limit of $\ep \to 0$.}. 
\vspace{4mm}

We  have successfully checked the following.  
\ice{369}
\begin{enumerate}
\item
All 369 p-integrals  
stop to depend on $\pi$ after being  rewritten  in terms of the proper generators \re{gensL7}
{\em provided} all terms proportional to
$\z_{12}$ are set to zero by  hand.

\item The  disappearance of $\pi$-dependence in the above point holds not only 
for  all terms with weight  less or equal  11 but also for all (rather numerous)  terms with the
transcendental weight 12 {\em and} 13.  

\item In fact, the database \cite{Panzer:2016snt} contains also many
  finite p-integrals with loop number  8.  Some of them depend
  on multiple  zeta values only.  If one discards in these integrals all
  contributions with the transcendental weight strictly larger than 13 then 
  there will remain exactly two non-vanishing  integrals. After  rewriting  the
  survivors  in terms of generators   \re{gensL7} and setting $\z_{12}$ to zero
  they also cease to depend on $\pi$!

\end{enumerate}

\ice{
This  is important as our hatted representation for the 7-loop case (eqs. (\ref{} --
\ref{}) is based on a rather small probe set of p-integrals it is
important 
  }

\section{Conclusion}

Using as input data essentially {\bf only} deep $\ep$ expansions  of the {\bf  4-loop} master
integrals \cite{Lee:2011jt} we have extended the hatted representation of
4-loop p-integrals of work \cite{Baikov:2010hf} to the 5-, 6- and 7-loop
families of p-integrals. At  5-loop level we successfully reproduced the 
results of \cite{Georgoudis:2018olj} which had been  obtained by a direct calculation  of 
a rather large subset of {\bf 5-loop} master p-integrals. 

We have derived a set of generic model-independent predictions for
$\pi$-dependent terms of RG-functions at 7 and 8 loops (at the latter case
only for terms with weight less or equal to  11). All available  7- and 8-loop
results are in agreement  with  our predictions.

Our results demonstrate a remarkable and somewhat mysterious (at least for us)
connection between $\ep$-expansions of the {\bf 4-loop} p-integrals and $D=4$
values of 5-, 6- and 7-loop {\em finite} p-integrals. \ice{(or, equivalently, with 6-,
7- and 8-loop periods).} Indeed, dealing {\em only} with 4-loop p-integrals we
have been able to get some non-trivial information about 5-, 6- and 7-loop
p-integrals. More precisely, we have found a set of proper transcendental
generators which form a $\pi$-free basis for every known 5-,6, and 7-loop
p-integrals provided that (i) the latter is expressible only in terms of multiple
zeta values and (ii) all terms (if any)  proportional to $\z_{12}$ are discarded.
 
It would be  interesting to see  what new information can be  extracted from
expanding 4-loop master p-integrals  to  even higher orders in $\ep$.

\acknowledgments 
%We are grateful to E. Panzer and V. Smirnov for useful discussions and  good advice.

We are very grateful to   V. Smirnov  for providing us with extension  of  the results  of \cite{Lee:2011jt}
to the transcendentality weight thirteen. 

We  thank  J. Gracey for his help in extracting 6-, 7- 
and 8-loop terms from generic results of
\cite{Gracey:1996he,Ciuchini:1999cv,Ciuchini:1999wy} and to O.~Schnetz
for providing us with his results for the 7-loop ADs in the $O(n)$
$\phi^4$ model.

The work of P.A.~Baikov is supported in part by the 
grant RFBR 17-02-00175A of the Russian Foundation for Basic Research.
The work by K. G. Chetykin was supported by 
the German Federal Ministry for Education and Research BMBF
through Grant  No. \ice{05H2015  and}05H15GUCC1 and
by the Deutsche Forschungsgemeinschaft through CH 1479/2-1.

%\mbib

\providecommand{\href}[2]{#2}\begingroup\raggedright\endgroup

\ed

\ed

9This is well-defined only conjecturally, because there could be additional relations between
MZVs which are not shared by motivic MZVs, making the replacement ζ (n) → ζ m (n) ambiguous.

^ TTP19-026 Transcendental structure of  multiloop    massless correlators and anomalous dimensions^
|**P.~A.~Baikov,  K. G. Chetyrkin** | 
| {{|PDF}} {{|PostScript}} [[|arXiv]]   |  
| |
\begin{thebibliography}{10}

\bibitem{Baikov:2018wgs}
P.~A. Baikov and K.~G. Chetyrkin, \emph{{The structure of generic anomalous
  dimensions and no-$\pi$ theorem for massless propagators}},
  \href{http://dx.doi.org/10.1007/JHEP06(2018)141}{\emph{JHEP} {\bfseries 06}
  (2018) 141}, [\href{https://arxiv.org/abs/1804.10088}{{\ttfamily
  1804.10088}}].

\bibitem{Schnetz:2016fhy}
O.~Schnetz, \emph{{Numbers and Functions in Quantum Field Theory}},
  \href{http://dx.doi.org/10.1103/PhysRevD.97.085018}{\emph{Phys. Rev.}
  {\bfseries D97} (2018) 085018},
  [\href{https://arxiv.org/abs/1606.08598}{{\ttfamily 1606.08598}}].

\bibitem{Gracey:2018ame}
J.~A. Gracey, \emph{{Large $N_f$ quantum field theory}},
  \href{http://dx.doi.org/10.1142/S0217751X18300326}{\emph{Int. J. Mod. Phys.}
  {\bfseries A33} (2019) 1830032},
  [\href{https://arxiv.org/abs/1812.05368}{{\ttfamily 1812.05368}}].

\bibitem{Gorishnii:1991vf}
S.~G. Gorishny, A.~L. Kataev and S.~A. Larin, \emph{The { ${\cal
  O}(\alpha_s^3)$} corrections to {$\sigma_{\rm tot}(e^+ e^- \to {\rm
  hadrons})$} and {$\sigma( {\tau} \to \nu_{\tau} + {\rm hadrons})$} in {
  QCD}}, {\emph{Phys. Lett.} {\bfseries B259} (1991) 144--150}.

\bibitem{MR360116}
R.~Ayoub, \emph{Euler and the zeta function},
  \href{http://dx.doi.org/10.2307/2319041}{\emph{Amer. Math. Monthly}
  {\bfseries 81} (1974) 1067--1086}.

\bibitem{Baikov:2010je}
P.~A. Baikov, K.~G. Chetyrkin and J.~H. K{\"u}hn, \emph{{Adler Function,
  Bjorken Sum Rule, and the Crewther Relation to Order $\alpha_s^4$ in a
  General Gauge Theory}},
  \href{http://dx.doi.org/10.1103/PhysRevLett.104.132004}{\emph{Phys. Rev.
  Lett.} {\bfseries 104} (2010) 132004},
  [\href{https://arxiv.org/abs/1001.3606}{{\ttfamily 1001.3606}}].

\bibitem{Baikov:2017ujl}
P.~A. Baikov, K.~G. Chetyrkin and J.~H. K{\"u}hn, \emph{{Five-loop fermion
  anomalous dimension for a general gauge group from four-loop massless
  propagators}}, \href{http://dx.doi.org/10.1007/JHEP04(2017)119}{\emph{JHEP}
  {\bfseries 04} (2017) 119},
  [\href{https://arxiv.org/abs/1702.01458}{{\ttfamily 1702.01458}}].

\bibitem{Chetyrkin:1980pr}
K.~G. Chetyrkin, A.~L. Kataev and F.~V. Tkachov, \emph{{New Approach to
  Evaluation of Multiloop Feynman Integrals: The Gegenbauer Polynomial x Space
  Technique}},
  \href{http://dx.doi.org/10.1016/0550-3213(80)90289-8}{\emph{Nucl. Phys.}
  {\bfseries B174} (1980) 345--377}.

\bibitem{Jamin:2017mul}
M.~Jamin and R.~Miravitllas, \emph{{Absence of even-integer $\zeta$-function
  values in Euclidean physical quantities in QCD}},
  \href{http://dx.doi.org/10.1016/j.physletb.2018.02.030}{\emph{Phys. Lett.}
  {\bfseries B779} (2018) 452--455},
  [\href{https://arxiv.org/abs/1711.00787}{{\ttfamily 1711.00787}}].

\bibitem{Davies:2017hyl}
J.~Davies and A.~Vogt, \emph{{Absence of $\pi^2$ terms in physical anomalous
  dimensions in DIS: Verification and resulting predictions}},
  \href{http://dx.doi.org/10.1016/j.physletb.2017.11.036}{\emph{Phys. Lett.}
  {\bfseries B776} (2018) 189--194},
  [\href{https://arxiv.org/abs/1711.05267}{{\ttfamily 1711.05267}}].

\bibitem{Chetyrkin:2017bjc}
K.~G. Chetyrkin, G.~Falcioni, F.~Herzog and J.~A.~M. Vermaseren,
  \emph{{Five-loop renormalisation of QCD in covariant gauges}},
  \href{http://dx.doi.org/10.1007/JHEP12(2017)006,
  10.1007/JHEP10(2017)179}{\emph{JHEP} {\bfseries 10} (2017) 179},
  [\href{https://arxiv.org/abs/1709.08541}{{\ttfamily 1709.08541}}].

\bibitem{Ruijl:2018poj}
B.~Ruijl, F.~Herzog, T.~Ueda, J.~A.~M. Vermaseren and A.~Vogt,
  \emph{{R*-operation and five-loop calculations}},
  \href{http://dx.doi.org/10.22323/1.290.0011}{\emph{PoS} {\bfseries
  RADCOR2017} (2018) 011}, [\href{https://arxiv.org/abs/1801.06084}{{\ttfamily
  1801.06084}}].

\bibitem{Herzog:2018kwj}
F.~Herzog, S.~Moch, B.~Ruijl, T.~Ueda, J.~A.~M. Vermaseren and A.~Vogt,
  \emph{{Five-loop contributions to low-N non-singlet anomalous dimensions in
  QCD}}, \href{http://dx.doi.org/10.1016/j.physletb.2019.01.060}{\emph{Phys.
  Lett.} {\bfseries B790} (2019) 436--443},
  [\href{https://arxiv.org/abs/1812.11818}{{\ttfamily 1812.11818}}].

\bibitem{Baikov:2016tgj}
P.~A. Baikov, K.~G. Chetyrkin and J.~H. K{\"u}hn, \emph{{Five-Loop Running of
  the QCD coupling constant}},
  \href{http://dx.doi.org/10.1103/PhysRevLett.118.082002}{\emph{Phys. Rev.
  Lett.} {\bfseries 118} (2017) 082002},
  [\href{https://arxiv.org/abs/1606.08659}{{\ttfamily 1606.08659}}].

\bibitem{Herzog:2017ohr}
F.~Herzog, B.~Ruijl, T.~Ueda, J.~A.~M. Vermaseren and A.~Vogt, \emph{{The
  five-loop beta function of Yang-Mills theory with fermions}},
  \href{https://arxiv.org/abs/1701.01404}{{\ttfamily 1701.01404}}.

\bibitem{Luthe:2017ttg}
T.~Luthe, A.~Maier, P.~Marquard and Y.~Schroder, \emph{{The five-loop Beta
  function for a general gauge group and anomalous dimensions beyond Feynman
  gauge}}, \href{http://dx.doi.org/10.1007/JHEP10(2017)166}{\emph{JHEP}
  {\bfseries 10} (2017) 166},
  [\href{https://arxiv.org/abs/1709.07718}{{\ttfamily 1709.07718}}].

\bibitem{Baikov:2010hf}
P.~A. Baikov and K.~G. Chetyrkin, \emph{{Four Loop Massless Propagators: An
  Algebraic Evaluation of All Master Integrals}},
  \href{http://dx.doi.org/10.1016/j.nuclphysb.2010.05.004}{\emph{Nucl. Phys.}
  {\bfseries B837} (2010) 186--220},
  [\href{https://arxiv.org/abs/1004.1153}{{\ttfamily 1004.1153}}].

\bibitem{Lee:2011jt}
R.~N. Lee, A.~V. Smirnov and V.~A. Smirnov, \emph{{Master Integrals for
  Four-Loop Massless Propagators up to Transcendentality Weight Twelve}},
  \href{http://dx.doi.org/10.1016/j.nuclphysb.2011.11.005}{\emph{Nucl. Phys.}
  {\bfseries B856} (2012) 95--110},
  [\href{https://arxiv.org/abs/1108.0732}{{\ttfamily 1108.0732}}].

\bibitem{Panzer:2013cha}
E.~Panzer, \emph{{On the analytic computation of massless propagators in
  dimensional regularization}},
  \href{http://dx.doi.org/10.1016/j.nuclphysb.2013.05.025}{\emph{Nucl. Phys.}
  {\bfseries B874} (2013) 567--593},
  [\href{https://arxiv.org/abs/1305.2161}{{\ttfamily 1305.2161}}].

\bibitem{Georgoudis:2018olj}
A.~Georgoudis, V.~Goncalves, E.~Panzer and R.~Pereira, \emph{{Five-loop
  massless propagator integrals}},
  \href{https://arxiv.org/abs/1802.00803}{{\ttfamily 1802.00803}}.

\bibitem{Brown:2015fyf}
F.~Brown, \emph{{Feynman amplitudes, coaction principle, and cosmic Galois
  group}}, \href{http://dx.doi.org/10.4310/CNTP.2017.v11.n3.a1}{\emph{Commun.
  Num. Theor. Phys.} {\bfseries 11} (2017) 453--556},
  [\href{https://arxiv.org/abs/1512.06409}{{\ttfamily 1512.06409}}].

\bibitem{Panzer:2016snt}
E.~Panzer and O.~Schnetz, \emph{{The Galois coaction on $\phi^4$ periods}},
  \href{http://dx.doi.org/10.4310/CNTP.2017.v11.n3.a3}{\emph{Commun. Num.
  Theor. Phys.} {\bfseries 11} (2017) 657--705},
  [\href{https://arxiv.org/abs/1603.04289}{{\ttfamily 1603.04289}}].

\bibitem{Broadhurst:1997kc}
D.~J. Broadhurst and D.~Kreimer, \emph{Association of multiple zeta values with
  positive knots via feynman diagrams up to 9 loops}, {\emph{Phys. Lett.}
  {\bfseries B393} (1997) 403--412},
  [\href{https://arxiv.org/abs/hep-th/9609128}{{\ttfamily hep-th/9609128}}].

\bibitem{Broadhurst2013}
D.~Broadhurst, \emph{Multiple Zeta Values and Modular Forms in Quantum Field
  Theory}, pp.~33--73.
\newblock Springer Vienna, Vienna, 2013.

\bibitem{Baikov:2018gap}
P.~A. Baikov and K.~G. Chetyrkin, \emph{{No-$\pi$ Theorem for Euclidean
  Massless Correlators}},
  \href{http://dx.doi.org/10.22323/1.303.0008}{\emph{PoS} {\bfseries LL2018}
  (2018) 008}, [\href{https://arxiv.org/abs/1808.00237}{{\ttfamily
  1808.00237}}].

\bibitem{Broadhurst:1999xk}
D.~J. Broadhurst, \emph{{Dimensionally continued multiloop gauge theory}},
  \href{https://arxiv.org/abs/hep-th/9909185}{{\ttfamily hep-th/9909185}}.

\bibitem{Blumlein:2009cf}
J.~Blumlein, D.~J. Broadhurst and J.~A.~M. Vermaseren, \emph{{The Multiple Zeta
  Value Data Mine}},
  \href{http://dx.doi.org/10.1016/j.cpc.2009.11.007}{\emph{Comput. Phys.
  Commun.} {\bfseries 181} (2010) 582--625},
  [\href{https://arxiv.org/abs/0907.2557}{{\ttfamily 0907.2557}}].

\bibitem{Broadhurst:1995km:improved}
D.~J. Broadhurst and D.~Kreimer, \emph{Knots and numbers in $\varphi^4$ theory
  to 7 loops and beyond}, {\emph{Int. J. Mod. Phys.} {\bfseries C6} (1995)
  519--524}, [\href{https://arxiv.org/abs/hep-ph/9504352}{{\ttfamily
  hep-ph/9504352}}].

\bibitem{Panzer:2015ida}
E.~Panzer, \emph{{Feynman integrals and hyperlogarithms}}.
\newblock PhD thesis, Humboldt U., Berlin, Inst. Math., 2015.
\newblock \href{https://arxiv.org/abs/1506.07243}{{\ttfamily 1506.07243}}.
\newblock 10.18452/17157.

\bibitem{Gracey:1996he}
J.~Gracey, \emph{{The QCD Beta function at ${\cal O}(1/N_f)$}},
  \href{http://dx.doi.org/10.1016/0370-2693(96)00105-0}{\emph{Phys.Lett.}
  {\bfseries B373} (1996) 178--184},
  [\href{https://arxiv.org/abs/hep-ph/9602214}{{\ttfamily hep-ph/9602214}}].

\bibitem{Ciuchini:1999cv}
M.~Ciuchini, S.~E. Derkachov, J.~Gracey and A.~Manashov, \emph{{Quark mass
  anomalous dimension at ${\cal O}(1/N_f^2)$ in QCD}},
  \href{http://dx.doi.org/10.1016/S0370-2693(99)00573-0}{\emph{Phys.Lett.}
  {\bfseries B458} (1999) 117--126},
  [\href{https://arxiv.org/abs/hep-ph/9903410}{{\ttfamily hep-ph/9903410}}].

\bibitem{Ciuchini:1999wy}
M.~Ciuchini, S.~E. Derkachov, J.~Gracey and A.~Manashov, \emph{{Computation of
  quark mass anomalous dimension at ${\cal O}(1/N_f^2)$ in quantum
  chromodynamics}},
  \href{http://dx.doi.org/10.1016/S0550-3213(00)00209-1}{\emph{Nucl.Phys.}
  {\bfseries B579} (2000) 56--100},
  [\href{https://arxiv.org/abs/hep-ph/9912221}{{\ttfamily hep-ph/9912221}}].

\bibitem{Vasiliev:1981yc}
A.~N. Vasiliev, {\relax Yu}.~M. Pismak and {\relax Yu}.~R. Khonkonen,
  \emph{{Simple Method of Calculating the Critical Indices in the 1/$N$
  Expansion}}, \href{http://dx.doi.org/10.1007/BF01030844}{\emph{Theor. Math.
  Phys.} {\bfseries 46} (1981) 104--113}.

\bibitem{Vasiliev:1981dg}
A.~N. Vasiliev, {\relax Yu}.~M. Pismak and {\relax Yu}.~R. Khonkonen,
  \emph{{1/$N$ Expansion: Calculation of the Exponents $\eta$ and $\nu$ in the
  Order 1/$N^2$ for Arbitrary Number of Dimensions}},
  \href{http://dx.doi.org/10.1007/BF01019296}{\emph{Theor. Math. Phys.}
  {\bfseries 47} (1981) 465--475}.

\bibitem{Vasiliev:1982dc}
A.~N. Vasiliev, {\relax Yu}.~M. Pismak and {\relax Yu}.~R. Khonkonen,
  \emph{{1/N Expansion: Calculation Of the Exponent $\eta$ in the Order $1/N^3$
  by the Conformal Bootstrap Method}},
  \href{http://dx.doi.org/10.1007/BF01015292}{\emph{Theor. Math. Phys.}
  {\bfseries 50} (1982) 127--134}.

\bibitem{Broadhurst:1996ur}
D.~J. Broadhurst, J.~A. Gracey and D.~Kreimer, \emph{{Beyond the triangle and
  uniqueness relations: Nonzeta counterterms at large N from positive knots}},
  \href{http://dx.doi.org/10.1007/s002880050500}{\emph{Z. Phys.} {\bfseries
  C75} (1997) 559--574},
  [\href{https://arxiv.org/abs/hep-th/9607174}{{\ttfamily hep-th/9607174}}].

\bibitem{Kotikov:2019bqo}
A.~V. Kotikov and S.~Teber, \emph{{On the Landau-Khalatnikov-Fradkin
  transformation and the mystery of even $\zeta$-values in Euclidean massless
  correlators}},  \href{https://arxiv.org/abs/1906.10930}{{\ttfamily
  1906.10930}}.

\end{thebibliography}
